\documentclass[journal]{IEEEtran}
\usepackage[utf8]{inputenc}
\usepackage[T1]{fontenc}
\usepackage{url}
\usepackage{ifthen}
\usepackage{cite}
\usepackage[cmex10]{amsmath}
\usepackage{stfloats}
\usepackage{graphicx}
\usepackage{epstopdf}
\usepackage{amsfonts}
\usepackage{enumitem}
\usepackage{amssymb}
\usepackage{amsfonts}
\usepackage{mathrsfs}
\usepackage{bm}
\usepackage{mdwmath}
\usepackage{subfigure}
\usepackage{mdwtab}
\usepackage{dsfont}
\usepackage{color}
\usepackage{verbatim}
\usepackage{array}
\usepackage{epstopdf}
\usepackage{multirow}
\usepackage{booktabs}
\usepackage{lineno}
\usepackage{amsmath}
\usepackage{makecell}
\usepackage{algorithmic}
\usepackage{pifont}
\usepackage{blindtext}
\usepackage{psfrag}
\usepackage{setspace}

\usepackage{flushend}
\usepackage[ruled,vlined,linesnumbered]{algorithm2e}

\allowdisplaybreaks

\newtheorem{Prob}{Problem}

\usepackage{titletoc}

\renewcommand\thesubsubsection{\alph{subsubsection}}

\begin{document}
\title{Quality of Experience Oriented Cross-layer Optimization for Real-time XR Video Transmission}


\author{Guangjin Pan, Shugong Xu, \IEEEmembership{Fellow, IEEE}, Shunqing Zhang, \IEEEmembership{Senior Member, IEEE},\\ Xiaojing Chen, \IEEEmembership{Member, IEEE} , and Yanzan Sun, \IEEEmembership{Member, IEEE} \\
% $^{\dagger}$
% School of Communication and Information Engineering, \\ Shanghai University, Shanghai 200444, China\\
% Email: \{guangjin\_pan, shugong, shunqing, jodiechen, yanzansun\}@shu.edu.cn\\
% \thanks{ Shugong Xu is the corresponding author.}
	\thanks{
		G. Pan, S. Xu, S. Zhang, X. Chen and Y. Sun are with School of Communication and Information Engineering, Shanghai University, Shanghai 200444, China. Emails: \{guangjin\_pan, shugong, shunqing, jodiechen, yanzansun\}@shu.edu.cn.

        Part of this work has been accepted by FCN-2023. This work was supported in part by the National Key R\&D Program of China under Grant 2022YFB2902005, the National High Quality Program grant TC220H07D, the National Natural Science Foundation of China (NSFC) under Grant 61871262, 62071284, and 61901251, the National Key R\&D Program of China grants 2022YFB2902000, Foshan Science and Technology Innovation Team Project grant FS0AAKJ919-4402-0060. The corresponding author is Shugong Xu.
        }

}

% \IEEEpubid{\begin{minipage}{\textwidth}\ \\[30pt] \centering
% 	Copyright © 2024 IEEE. Personal use of this material is permitted. However, permission to use this material for any other purposes must be obtained from the IEEE by sending an email to pubs-permissions@ieee.org.
% \end{minipage}}

\markboth{IEEE TRANSACTIONS ON CIRCUITS AND SYSTEMS FOR VIDEO TECHNOLOGY, ~Vol.~XX, No.~XX, Feb~2024}%
{Shell \MakeLowercase{\textit{et al.}}: Bare Demo of IEEEtran.cls for IEEE Transactions on Magnetics Journals}

\IEEEtitleabstractindextext{%
\begin{abstract}
Extended reality (XR) is one of the most important applications of beyond 5G and 6G networks. Real-time XR video transmission presents challenges in terms of data rate and delay. In particular, the frame-by-frame transmission mode of XR video makes real-time XR video very sensitive to dynamic network environments. To improve the users' quality of experience (QoE), we design a cross-layer transmission framework for real-time XR video. The proposed framework allows the simple information exchange between the base station (BS) and the XR server, which assists in adaptive bitrate and wireless resource scheduling. We utilize the cross-layer information to formulate the problem of maximizing user QoE by finding the optimal scheduling and bitrate adjustment strategies. To address the issue of mismatched time scales between two strategies, we decouple the original problem and solve them individually using a multi-agent-based approach. Specifically, we propose the multi-step Deep Q-network (MS-DQN) algorithm to obtain a frame-priority-based wireless resource scheduling strategy and then propose the Transformer-based Proximal Policy Optimization (TPPO) algorithm for video bitrate adaptation. The experimental results show that the TPPO+MS-DQN algorithm proposed in this study can improve the QoE by 3.6\% to 37.8\%. More specifically, the proposed MS-DQN algorithm enhances the transmission quality by 49.9\%-80.2\%.

\end{abstract}
}

\maketitle

\IEEEdisplaynontitleabstractindextext

\IEEEpeerreviewmaketitle

\section{Introduction}
\IEEEPARstart{R}{ecently}, Metaverse \cite{Metaverse} sparks people's imagination about immersive experiences. As a result, extended reality (XR) becomes one of the most critical 5G media applications under deliberation, because XR can provide people with immersive experiences. XR encompasses various types of reality, including Augmented Reality (AR), Virtual Reality (VR), Mixed Reality (MR) and Cloud Gaming (CG) \cite{cite:TS26928,cite:XR}, all of which require low-latency transmission and interaction. It poses new challenges for beyond 5G and 6G networks.

The use of XR in amalgamating virtual and real worlds presents users with an immersive experience that requires significant computing power. Hence, XR services are generally rendered on cloud or edge servers and then transmitted to the XR clients for playback. XR services fall under the category of Real-Time Broadband Communication (RTBC) services, which comprise enhanced mobile broadband (eMBB) communication and ultra-reliable low latency communication (URLLC) requirements. To ensure a seamless, high-quality XR video experience, the system must fulfill the standards of low latency and high data rate. Unfortunately, the existing adaptive real-time XR video streaming systems over wireless networks pose two main challenges.
\begin{itemize}
  \item{\textit{The challenge of bitrate control for XR servers.}} For XR servers, the challenge is to adjust the bitrate rapidly and accurately so as to adapt to the network environment. In traditional chunk-based video adaptive wireless streaming systems, users combine historical bandwidth and buffer status to adjust the bitrate adaptively \cite{cite:Pensieve}. Updates to bandwidth estimation and buffer status typically take seconds. However, second-level bandwidth estimation cannot meet frame-level transmission requirements in real-time XR systems. Due to the frame-by-frame transmission and playback mode of the real-time XR video, even slight network jitter can impact the performance of XR transmission. To overcome this challenge, the bitrate of XR videos can be adjusted based on the user feedback information, such as loss rate and packet arrival delay gradient \cite{cite:GCC}. However, this information can only indirectly reflect the network status, and the system lacks lower-layer network observations for bitrate adjustment.
  \item{\textit{The Challenge of resource scheduling for the base station (BS)}.} 5G has larger transmission bandwidth and lower latency to meet the requirements of eMBB and URLLC scenarios. The real-time XR transmission requires both low latency and high data rates, which poses new challenges for wireless systems. Additionally, different XR video frames may have varying levels of importance. For example, in the Group of Picture (GOP)-based XR video encoding model \cite{cite:TS38838}, I-frames are usually more important than P-frames. Therefore, if wireless resources are limited, the scheduler should prioritize transmitting important information. However, traditional scheduling policies based on traffic flow priority, such as \cite{cite:M-EDF-PF} and \cite{cite:priority-DDPG}, can not handle this problem well. Meanwhile, the dynamic wireless environment and resources make real-time XR video transmission more difficult.
\end{itemize}

These challenges motivate our study on a new cross-layer transmission architecture for real-time XR video. We consider a real-time XR oriented wireless transmission system. As shown in Fig. \ref{fig:architecture}, the edge XR server performs real-time video encoding and transmits each encoded video frame to BS with each frame consisting of a set of video packets. Subsequently, the BS schedules the video packets in the queue and delivers them to the respective XR users, who then decode and play the video. The contributions of our study are listed below,
\begin{itemize}
  \item{\textit{A cross-layer transmission framework for real-time XR video.}} We propose a framework for real-time XR video transmission. The proposed framework enables the XR server and BS to assist each other in improving the quality of experience (QoE) for XR videos. Based on the proposed framework, we model the problem as a cross-layer optimization problem between the Medium Access Control (MAC) and application layers, intending to improve the QoE of real-time XR video. In order to address the problem of mismatch between the wireless resource allocation and adaptive bitrate at two time-scales, we decouple the problem into two problems and then solve each separately.
  \item{\textit{A frame-priority-based wireless resource scheduling scheme at the MAC layer.}} At the MAC layer, we propose a frame-priority based scheduling scheme to maximize the frame success rate. The resource allocation for each time slot involves a highly complex and large action space, as it requires the allocation of a variable number of resource blocks (RBs) for each video frame. To solve this problem, we propose a multi-step Deep Q-network (MS-DQN) algorithm to reduce the action space, allowing the agent to make multi-step decisions within each time slot to allocate wireless resources.
  \item{\textit{A TPPO-based adaptive bitrate scheme at the application layer.}} At the application layer, due to the XR server's inability to obtain global information, the problem is modelled as a partially observable Markov decision process (POMDP). Therefore, we need a method that can extract the underlying relationships between video bitrate selection and communication modes from historical observation sequences to choose the appropriate video bitrate. We introduce a Transformer-based Proximal Policy Optimization (TPPO) algorithm to extract semantic information from the historical environment and intelligently select the optimal video bitrate to improve the QoE of real-time XR video.   
\end{itemize}

Compared with the baseline algorithms Proportional Fair (PF) \cite{PF}, PF-I \cite{cite:frameintegrated}, and Deep Deterministic Policy Gradient (DDPG) \cite{cite:priority-DDPG}, the proposed MS-DQN explores and learns the optimal scheduling strategy through interaction with the environment. At the same time, MS-DQN reduces the action space of each step decision by decomposing the single-step decision process into a multi-step decision, thus reducing the learning difficulty. In comparison with baseline algorithms DDPG \cite{cite:DDPGABR} and Proximal Policy Optimization (PPO) \cite{cite:PPOABR}, the proposed bitrate adjustment algorithm TPPO combines the advantages of the PPO algorithm and transformer to extract semantic information from the historical environment. The numerical results show that our proposed TPPO+MS-DQN algorithm can outperform the baseline algorithms. The MS-DQN-based algorithm can increase the transmission quality by 49.9\%-80.2\%, and the proposed TPPO+MS-DQN algorithm can improve QoE by 3.6\%-37.8\%.

The rest of this paper is organized as follows. Sec. \ref{Sec:RelatedWork} provides a detailed overview of related works. In Sec. \ref{Sec:SystemModel}, we introduce system models , and in Sec. \ref{Sec:SystemModel}, we formulate the problem. In Sec. \ref{Sec:MSDQN}, we describe the proposed MS-DQN algorithm. Sec. \ref{Sec:TPPO} gives the proposed TPPO algorithm. Numerical results and analysis are presented in Sec. \ref{Sec:Result}. Finally, the paper is concluded in Section \ref{Sec:Conclusion}.

\section{Related Work} \label{Sec:RelatedWork}

Several related works have been developed on wireless scheduling, as discussed in \cite{PF,cite:exprule,cite:M-EDF-PF,cite:priority-DDPG,cite:RPPO,cite:frameintegrated}. The proportional fair scheduler was analyzed in \cite{PF} to obtain the cell throughput, while scheduling algorithms like Exponential Rule (EXP-RULE) \cite{cite:exprule} and Modified Earliest Deadline First and Proportion Fair (M-EDF-PF) \cite{cite:M-EDF-PF} were proposed to support real-time traffics. In \cite{cite:priority-DDPG}, the DDPG algorithm was utilized to minimize queuing delay experienced by users. Additionally, a delay-oriented scheduling algorithm based on partially observable Markov decision process was proposed in \cite{cite:RPPO} to reduce the tail delay and average delay. However, these works do not consider the traffic characteristics of XR. \cite{cite:frameintegrated} focused on frame-level integrated transmission to enhance the frame success rate.

Adaptive bitrate algorithms have been extensively researched for chunk-based video-on-demand transmission. \cite{cite:ControlTheoretic, cite:Pensieve, cite:BOLA, cite:Fleet,cite:Panda,cite:MUABR,cite:MULyapunov,cite:PPOABR} use bandwidth prediction, video buffer size, or both to adjust bitrates of video chunks to improve the QoE. \cite{cite:ControlTheoretic} uses a control-theoretic method for dynamic adaptive bitrate streaming, and proposes a model predictive control algorithm. Pensieve \cite{cite:Pensieve} uses the Deep Reinforcement Learning (DRL) algorithm to learn and make adaptive bitrate (ABR) decisions solely by observing the result performance of past decisions. BOLA \cite{cite:BOLA} uses Lyapunov optimization to minimize rebuffering and maximize video quality. Fleet \cite{cite:Fleet} considers the idle period of the Live video streaming, and designs a HyperText Transfer Protocol (HTTP) chunk-level measurement algorithm and a stochastic model predictive control (MPC) controller for bitrate control. These works only consider the QoE of a single user, and cannot be applicable to scenarios where multiple video streams compete for network resources. The authors in \cite{cite:Panda} believe when multiple clients compete at a network bottleneck, it is difficult for a client to correctly perceive its fair-share bandwidth. They propose the PANDA \cite{cite:Panda} algorithm to probe the fair-share bandwidth. The authors in \cite{cite:MUABR} propose the MUABR algorithm based on multi-agent deep reinforcement learning, which combines ABR algorithm with bandwidth allocation strategy to optimize the overall QoE and can improve viewing experience of each user as much as possible. The authors in \cite{cite:MULyapunov} employ Lyapunov optimization techniques for bandwidth allocation among multiple users. In \cite{cite:PPOABR}, the authors further enhanced the average QoE of users using a PPO-based ABR algorithm. The aforementioned schemes address the problem of bitrate adaptation through the perspective of end-to-end optimization but do not consider the role of wireless systems.

In wireless video transmission systems, reasonable resource allocation in the BS helps improve the quality of video transmission and ensures fairness among users \cite{cite:MIMOvideo}. Correspondingly, the lower layer information in the BS is also helpful in adjusting the video bitrate of users \cite{cite:DDQNABR}. Therefore, a large amount of research has focused on the cross-layer optimization in wireless video transmission systems \cite{cite:OFDMDASH,cite:MIMOvideo,cite:DDQNABR,cite:NOMAVIDEO,cite:TranscodeABR,cite:DDPGABR}. \cite{cite:DDQNABR} combines higher and lower layers of information and proposes a QoE-oriented adaptive video streaming algorithm based on dueling DQN. In multiple input multiple output (MIMO) systems, authors in \cite{cite:MIMOvideo} jointly optimize the beamforming vector and video bitrate, and propose an optimization-based beamforming scheme and a DRL-based bitrate adaptation scheme. In orthogonal frequency division multiple (OFDM) systems, \cite{cite:OFDMDASH} jointly optimize communication resources, transmit power, and the video bitrate. They derive an online optimization problem to approximate the offline problem, and solve the online problem by decomposing it into wireless resource allocation problem and bitrate adaptation problem. In \cite{cite:NOMAVIDEO}, the authors propose an non-orthgonal multiple access (NOMA) based uncoded multiuser video transmission system. They optimize the allocation of power and channel resources to guarantee a high-quality linear video delivery. The authors in \cite{cite:TranscodeABR} aim to design an online joint transcoding and transmission resource allocation algorithm to maximize the users' QoE. In \cite{cite:DDPGABR}, the authors jointly design caching, computing, and bitrate adaption to improve user's QoE. However, the above works only consider the transmission and bitrate adaption of chunk-based video and cannot be applied to the frame-based real-time XR video transmission. In \cite{cite:RD1}, the authors consider the uplink multi-user video transmission in an OFDM system. They utilize channel state information and rate distortion (RD) information for subcarrier allocation and power allocation, aiming to optimize the sum of distortions at each time slot. Paper \cite{cite:RD2} further investigates the cross-layer resource allocation problem in a multi-antenna BF-NOMA-OFDMA system, with the goal of reducing the sum of MSE distortions in video transmission. The authors in \cite{cite:RD3} consider multi-user video communication on the uplink OFDMA system and a cross-layer algorithm is proposed for joint bit allocation, packet scheduling and wireless resource assignment are proposed to minimize the end-to-end expected video distortion. However, these existing works \cite{cite:RD1,cite:RD2,cite:RD3} do not consider the problem of mismatch between bitrate adaption and resource allocation at two time-scales. In addition, these studies model the quality of videos using the RD function at the level of GoP, which cannot be applied to real-time XR video transmission.

For real-time XR video transmission, Google propose the Google Congestion Control (GCC) congestion control algorithm for the Web real-time communication (WebRTC) system \cite{cite:GCC}. \cite{cite:GCCPlus} uses reinforcement learning and federated learning to optimize real-time video transmission. However, these end-to-end adaptive bitrate methods do not consider the assistance of wireless systems for video transmission. In \cite{cite:FECvideo}, 
a cross-layer resource allocation technique is proposed for wireless video transmission along with each forward error correction (FEC) block. The authors in \cite{cite:NOMA} consider a joint video packet assignment, power control and user scheduling problem to maximize the real-time video transmission quality for NOMA networks. However, they do not consider the bitrate adaptation.

\section{System Model} \label{Sec:SystemModel}

\begin{figure*}[tb]
\centering 
\includegraphics[height=3.3in,width=5.4in]{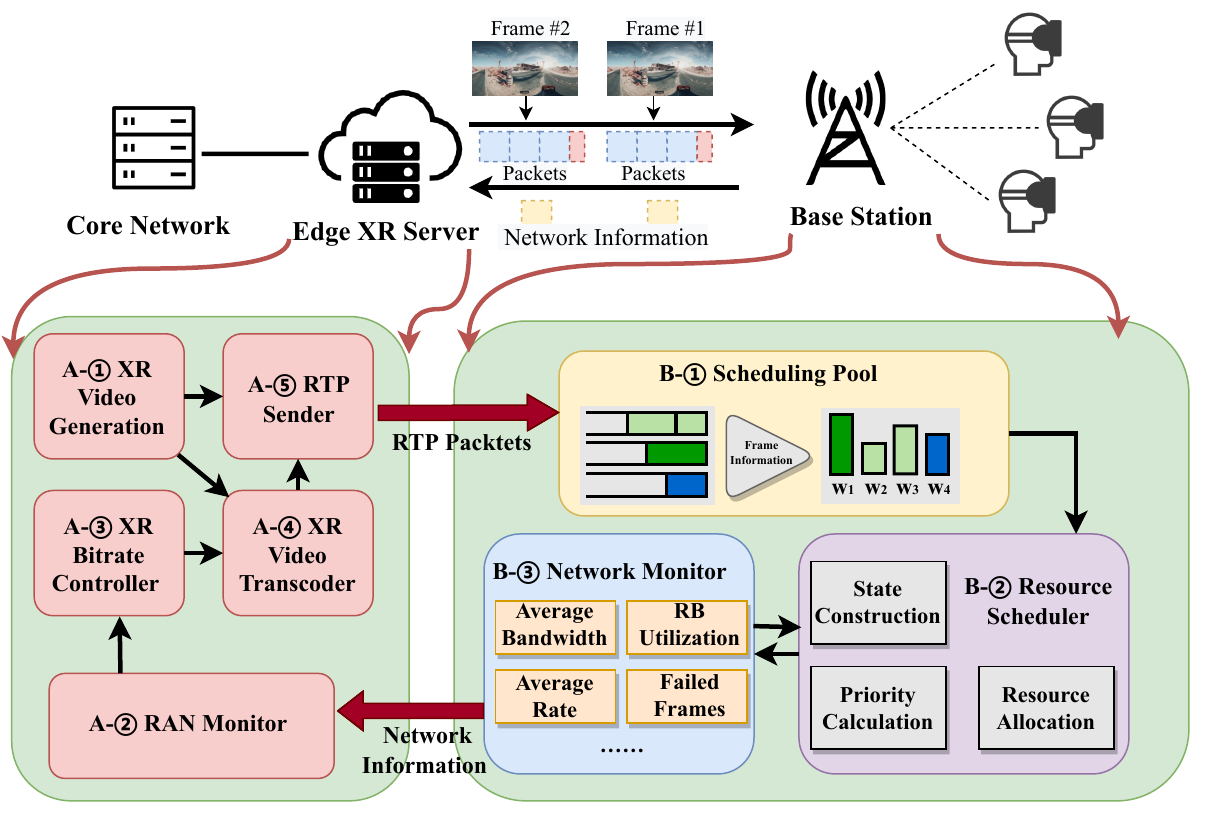}
\caption{XR transmission system framework. Real-time XR video is rendered on edge XR servers and then transmitted to clients for playback. BS uses frame information to schedule packets and improve transmission quality.}
\label{fig:architecture} 
\vspace{-2mm}
\end{figure*}

In this section, we consider an XR video 5G transmission system with one edge XR server, one BS, and $N$ XR users denoted by the set $\mathcal{N} = \{ 1, 2, \cdots, N \}$. The time is assumed to be discretized into several slots $ \mathcal{T}=\{0,1,\cdots,t,\cdots\}$. The length of one slot $\Delta t$ is equal to the time length of one transmission time interval (TTI) of the 5G system. Therefore, the BS allocates RBs once in each time slot $t$.

Fig. \ref{fig:architecture} shows our proposed framework for edge XR streaming. In this framework, the BS schedules XR video packets using a frame-priority-based scheduling scheme based on available frame information. The edge XR server monitors the network state of the BS to adjust the video bitrate dynamically. Specifically, in the edge XR server, the bitrate adjustment module includes A-\ding{172} Video Generator, A-\ding{173} Radio Access Network (RAN) Monitor, A-\ding{174} XR Bitrate Controller, A-\ding{175} XR Video Transcoder, and A-\ding{176} Real-Time Transport Protocol (RTP) Sender. The A-\ding{172} Video Generator generates XR videos in real time. The A-\ding{173} RAN Monitor collects the network information from the BS, including average rate, bandwidth, frame transmission success rate, and RB Utilization. The \ding{174} XR Bitrate Controller selects an appropriate bitrate based on the collected network information, and the A-\ding{175} XR Video Transcoder encodes the video stream accordingly. The A-\ding{176} RTP sender then packets the video and marks RTP header information before transmitting it to the BS via several IP packets. On the BS, the frame-based scheduling pool is constructed by the B-\ding{172} Scheduling Pool to identify and schedule packets according to the frame information from the RTP header. The B-\ding{173} Resource Scheduler in turn schedules packets based on network information. The B-\ding{174} Network Monitor periodically sends network information to the XR server to support dynamic bitrate adjustment.

\subsection{XR Traffic Model}

Considering the inherent timescale mismatch between the MAC layer and the application layer, we denote $T_b\Delta t$ as the duration of the bitrate adjustment period, and define that the bitrate adjusts in time slot set $\mathcal{T}_b =\{0,T_b,2T_b,\cdots,t_bT_b,\cdots\} \subset \mathcal{T}$. For convenience, we define $t_b$ as the bitrate adjustment period in time slot $t_bT_b$.

In the XR server, video frames are generated at regular intervals based on a predefined frame rate. One encoding scheme is GOP-based, where I-frames are transmitted periodically every $K$ frames, and the remaining frames are P-frames \cite{cite:TS38838}. Let $\mathcal{K}_n$ denote the frame set of user $n$. $\mathcal{K}_{n,t_b} \subset \mathcal{K}_n$ is the set of frames of user $n$ within the bitrate adjustment period $t_b$. The frame rate of the user's XR video is denoted as $f_n$, and $\alpha$ represents the average ratio of size between one I-frame and one P-frame \cite{cite:TS38838}. In the time slot $ t_bT_b \le t <(t_b+1)T_b$, the video bitrate of user $n$ is $R_{n,t_b}$, and the frame size of the $k$-th ($k \in  \mathcal{K}_{n,t_b}$) video frame arriving is represented by $r_{n,k}$. As reported in \cite{cite:TS38838}, the average size of $r_{n,k}$ is,
\begin{eqnarray}\label{equ:framesize}  
\mathbb{E}[r_{n,k}]= \!  \left\{
{\begin{aligned}
\frac{R_{n,t_b}}{f_n} \cdot \frac{K\alpha}{K-1+\alpha}  , \ \mathop{\textrm{if}} d_{n,k}=1,\\
\frac{R_{n,t_b}}{f_n} \cdot \frac{K}{K-1+\alpha} , \  \mathop{\textrm{if}} d_{n,k}=0  \nonumber,
\end{aligned}}  \right.
\end{eqnarray}
where $d_{n,k}$ is used to indicate the type of frame, with a value of either 0 or 1. Specifically, if $d_{n,k}=1$, the frame is an I-frame, while a value of 0 corresponds to a P-frame.

The $k$-th frame's arrival time of user $n$ can be expressed as,
\begin{eqnarray}\label{equ2-A-1}  
t_{n,k}= t_{n,0}+\lfloor \frac{\frac{k-1}{f_n}+t_{jitter}}{\Delta t} \rfloor,
\end{eqnarray}
where $t_{n,0}$ is the initial arrival time, $f_n$ is the frame rate, and $t_{jitter}$ is the arrival jitter time due to the codec and transmission. The value of $t_{jitter}$ is a random variable that follows a certain distribution.

In addition, considering the different types of frames and encoding methods, different XR video frames have varying levels of importance. For example, I-frames contain all the information of a video frame and can be decoded independently, while P-frames require the information from the previous frame for decoding. Therefore, compared to P-frames, I-frames have more data and are more important. For each user $n$, we define $w_{n,k}$ as the importance weight for frame $k$ \footnote{The optimal design of the importance weight depends on the type of video encoder and the content of the video. In this paper, we mainly discuss how to design scheduling algorithms to optimize real-time XR video transmission in the presence of frame importance weight. The optimization of importance weights is not our primary focus in this discussion.}.

\subsection{Channel and Scheduling Models} 
Considering there are $N^{rb}_t$ available RBs in time slot $t$, for each RB, the achievable data rate of user $n$ at the $t$-th slot is given by,
\begin{eqnarray}\label{equ2-B-1}  
{c}_{n,t}= \Delta{t} B_w \log(1+\frac{p_n h_{n,t}}{B_w\sigma^2}), \ \mathop{\textrm{bits/s}} 
\end{eqnarray}
where $p_n$ is the transmit power of user $n$, $h_{n,t}$ is the channel gain of user $n$ in time slot $t$,  $B_w$ is the bandwidth of each RB, $\sigma^2$ is the Gaussian noise power spectral density. ${c}_{n,t}$ represents the amount of data that each RB can transmit.

In the XR video transmission task, it is essential to fully transmit packets belonging to the same frame before the frame delay budget (FDB) $t^{\star}$ \cite{cite:frameintegrated}. Let $\widetilde{r}_{n,k,t}$ denote the remaining data size of frame $k$ waiting for transmission in time slot $t$. It can be given by,
\begin{eqnarray}\label{equ2-B-2}
\widetilde{r}_{n,k,t} = r_{n,k} - \sum_{\tau=t_{n,k}}^{t-1} N_{n,k,\tau}^{rb} {c}_{n,\tau},
\end{eqnarray}
where $r_{n,k}$ represents the total size of the frame, $N_{n,k,t}^{rb}$ denotes the number of RBs allocated to user $n$ for frame $k$ in time slot $t$, and ${c}_{n,\tau}$ represents the achievable data rate of user $n$ in RBs at time slot $\tau$, as calculated in Equation~\eqref{equ2-B-1}.

Let $x_{n,k}$ represent the indicator of whether the video frame is successfully transmitted. Therefore, a frame can be considered as successfully transmitted if the condition
\begin{eqnarray}\label{equ2-B-3}
x_{n,k} = \mathbb{I}(\widetilde{r}_{n,k,t_{n,k}+t^{\star}} \le 0) = 1
\end{eqnarray}
is satisfied, which indicates that the allocated wireless resources are capable of transmitting a larger amount of data than the size of the video frames. Otherwise, $x_{n,k} =0$.

To enhance the transmission capability of XR in wireless networks, we propose that the BS extracts some frame information from the real-time transport protocol (RTP) header\footnote{Although the current RTP standard \cite{cite:RTP} does not support this assumption, we believe that if the BS can obtain some frame information, the transmission quality of XR video can be effectively improved.}. By utilizing this information, the BS can sort the packets waiting to be transmitted and schedule them accordingly. Specifically, the set of frames waiting to be transmitted during time slot $t$ is denoted by $\mathcal{K}_{t}= \{(n,k)|n \in \mathcal{N}, k \in \mathcal{K}_{n,t} \}$, where $\mathcal{K}_{n,t}=\{k|k\in \mathcal{K}_n,0 \le t-t_{n,k} < t^{\star}, \widetilde{r}_{n,k,t}>0\}$ represents the set of frames for user $n$ that are waiting to be transmitted in time slot $t$.

The BS can perform scheduling based on the priority of frames, which is determined by the set of frames received. Let $p_{n,k,t}$ denote the priority of user $n$ for frame $k$ in time slot $t$. $N_{n,k,t}^{rb}$ is the number of RBs allocated to each video frame based on frame priority. The resource allocation policy can be expressed as,
\begin{eqnarray}\label{equ3-A-2}  
N_{n,k,t}^{rb}=\!  \left\{
{\begin{aligned}
\lceil \frac{\widetilde{r}_{n,k,t}}{{c}_{n,t}} \rceil   ,   \mathop{\textrm{if}} \widetilde{r}_{n,k,t}<{c}_{n,t}(N_{t}^{rb}- \! \! \! \! \! \! \! \! \! \! \! \sum_{(n',k')\in \mathcal{K}_{n,k,t}} \! \! \! \!  \! \! \! \! \! \! \!  N_{n',k',t}^{rb}),\\
N_{t}^{rb}- \! \! \! \! \! \! \! \! \! \! \! \sum_{(n',k')\in \mathcal{K}_{n,k,t}} \! \! \! \! \! \! \! \! \! N_{n',k',t}^{rb} \qquad \qquad \quad \ \ , \mathop{\textrm{otherwise}} ,
\end{aligned}}  \right. \label{equ:allocation}
\end{eqnarray}
where $\mathcal{K}_{n,k,t}=\{(n',k') | (n',k') \in \mathcal{K}_t, \ p_{n,k,t}<p_{n',k',t}\}$ is the set of frames with a priority higher than the frame $(n,k)$. The resource allocation policy indicates that, given the priority of video frames, the resource scheduler will prioritize allocating sufficient wireless resources to higher-priority video frames.

The traditional flow-priority-based scheduling policy primarily considers the user's queue lengths (to ensure fairness) and the channel state information (to optimize throughput) \cite{cite:RPPO}. In the frame-priority-based scheduler scheduler, the scheduler utilizes frame information and channel state information for scheduling. The priority is calculated by,
\begin{eqnarray}
\bm{p}_{t}={\pi}^p(\bm{{r}}_{t},\bm{\widetilde{r}}_{t},\bm{\tau}_t,\bm{w}_t,\bm{h}_t), \label{equ:priority}
\end{eqnarray}
where $\bm{p}_{t}\triangleq \{{p}_{n,k,t},\forall (n,k) \in \mathcal{K}_{t}\}$, $\bm{{r}}_t\triangleq\{{r}_{n,k},\forall (n,k) \in \mathcal{K}_{t}\}$, $\bm{\widetilde{r}}_t\triangleq\{\widetilde{r}_{n,k,t},\forall (n,k) \in \mathcal{K}_{t}\}$, $\bm{\tau}_t\triangleq\{\tau_{n,k,t},\forall (n,k) \in \mathcal{K}_{t}\}$, $\bm{w}_t\triangleq\{w_{n,k},\forall (n,k) \in \mathcal{K}_{t}\}$,  and $\bm{h}_t\triangleq\{h_{n,t},\forall (n) \in \mathcal{N}\}$. Specifically, $\tau_{n,k,t}=t^{\star}+t_{n,k}-t$ is the remaining FDB (RFDB). ${\pi}^p(\cdot)$ represents the scheduler policy based on video frame information and channel information.

\subsection{Adaptive XR Video Streaming Model}

As mentioned earlier, the XR server periodically gathers network information and performs XR video bitrate adaptation. Using the A-\ding{174} XR Bitrate Controller module described above, the XR Bitrate Controller utilizes historical observations of network information and selected bitrates to make decisions on video bitrate selection. Specifically, at the beginning of each bitrate adjustment period, the RAN Monitor samples network information $\bm{o}_{t_{b}}$ from the BS. The XR Bitrate Controller then uses the historical network information observation sequence $\bm{o}_{t_{b}},\cdots,\bm{o}_{t_{b}-t_o}$ and historical bitrate selection results $R_{1,t_b-1},\cdots,R_{1,t_b-t_o},\cdots,R_{N,t_{b}-1},\cdots, R_{N,t_{b}-t_o}$ to determine the video bitrate $R_{1,t_{b}}$ for that adjustment period. $t_o$ represents the length of the observation window. The input to the XR Bitrate Controller can be represented as $O_{t_{b}}=\{R_{1,t_b-1},\cdots, R_{1,t_b-t_o},\cdots, R_{N,t_{b}-1},\cdots, R_{N,t_{b}-t_o},\bm{o}_{t_{b}}, \cdots, \\ 
\bm{o}_{t_{b}-t_o}\}$. Therefore, the adaptive bitrate policy of the XR Bitrate Controller can be expressed as
\begin{eqnarray}
R_{n,t_b}={\pi}^b(O_{t_{b}}),
\label{equ:bitrate}
\end{eqnarray}
where ${\pi}^b(O_{t_{b}})$ denotes the adaptive bitrate strategy deployed for the XR video.

In chunk-based adaptive video streaming, the QoE of a user is typically defined in terms of video quality, quality variation, and rebuffering time \cite{cite:Pensieve}. However, the chunk-based QoE definition cannot directly apply to real-time XR video transmission systems. Inspired by \cite{cite:Pensieve}, we use a frame-based QoE expression to represent the QoE of adaptive real-time XR video streams. We redefine the QoE of real-time XR video into two parts, i.e., video playback quality $\mathcal{Q}_{n,t_b}^p $ and video transmission quality $\mathcal{Q}_{n,t_b}^t$. The video playback quality is mainly related to the video bitrate, including video quality and quality variations, and the quality of video transmission is related to whether the video frames are successfully transmitted, where
\begin{eqnarray}
\mathcal{Q}_{n,t_b}^p \! \! \! \! &=& \! \! \! \! \beta_1\frac{R_{n,t_b}|\mathcal{K}_{n,t_b}|}{f_n}-\beta_2|R_{n,t_b}-R_{n,t_{b-1}}| ,\label{equ:qoe_p}\\
\mathcal{Q}_{n,t_b}^t \! \! \! \! &=& \! \! \! \!-\sum_{k \in \widetilde{\mathcal{K}}_{n,t_b}} w_{n,k}.\label{equ:qoe_t}
\end{eqnarray}
$\beta_1$ and $\beta_2$ are the weight factors, and $\widetilde{\mathcal{K}}_{n,t_b}$ is the set of frames that failed to transmit within the FDB. The QoE mainly comprises the following three components,
\begin{itemize}
  \item{\textit{Video Quality}}. We define the QoE for video quality as the product of the average frame size and the number of frames transmitted within a bitrate adjustment period, representing the total size of the video sent. It is represented by the first item of \eqref{equ:qoe_p}.
  \item{\textit{Quality Variations.}} 
  Similar to \cite{cite:Pensieve}, the quality variations indicate the magnitude of video quality variation within adjacent bitrate adjustment periods, used to measure the stability of the video bitrate. It is represented by the second item of \eqref{equ:qoe_p}.
  \item{\textit{Weight Sum of Failed Frames}.} Transmission failures of frames lead to considerable QoE losses, impacting the user's overall experience negatively. The QoE of transmission is defined as the sum of the importance weight for failed frames during a bitrate adjustment period. We can express the transmission QoE using \eqref{equ:qoe_t}.

\end{itemize}

\section{Problem Formulation} \label{Sec3}
In this paper, our goal is to maximize the QoE of XR video applications. Based on the proposed XR transmission system framework, we aim to find the optimal video bitrate adjustment scheme ${\pi^{b}}(\cdot)$ and wireless resource allocation scheme ${{\pi}^p}(\cdot)$. 

\begin{Prob}
[Original Problem] The original problem can be formulated as the following optimization problem, \label{Prob:original}
\begin{eqnarray}
\mathop{\textrm{maximize}}_{{{\pi}^p}(\cdot),{{\pi}^b}(\cdot)} && \! \! \! \! \sum_{n \in \mathcal{N} }\mathbb{E}[\alpha_p\log \mathcal{Q}_{n,t_b}^p  + \alpha_t \mathcal{Q}_{n,t_b}^t]  \label{equ:originalproblem}  \\
\mathop{\textrm{s.t.}} 
&& \! \! \! \! \sum_{(n,k) \in \mathcal{K}_t}N_{n,k,t}^{rb} \le N_t, \ \forall{t}, \label{equ:RBconstraint}\\
&& \! \! \! \! R_{min} \le R_{n,t_b} \le R_{max}, \ \forall{t_b}, \label{equ:Biterateconstraint}\\
&& \! \! \! \! \eqref{equ:allocation},\eqref{equ:priority},\eqref{equ:bitrate},\eqref{equ:qoe_p},\eqref{equ:qoe_t}, \nonumber
\end{eqnarray}
\end{Prob}
where $\alpha_p$ and $\alpha_t$ are weights. $R_{max}$ and $R_{min}$ represent the maximum and minimum XR video bitrate, respectively. In \eqref{equ:originalproblem}, the objective QoE for XR video transmission consists of the video playback quality and the video transmission quality. The video playback quality uses a logarithmic function to improve fairness among users' bitrates. Meanwhile, we use the parameters $\alpha_p$ and $\alpha_t$ to balance the video playback quality and the video transmission quality. \eqref{equ:RBconstraint} constrains the number of RBs in each time slot, and \eqref{equ:Biterateconstraint} constrains the range of bitrate selection.
\color{black}

The above problem is generally difficult to solve due to the following reasons. Firstly, the problem couples two time-scales. The wireless resource allocation task is performed each slot $t$ and the bitrate selection task is performed each bitrate adjustment period $t_b$. Secondly, as a long-term optimization problem, the uncertainty of wireless channel state, available network resources, arrival time of frames, frame sizes poses a challenge to effective problem solving. Thirdly, for wireless resource allocation task, it is difficult to obtain an optimal closed-form expression for the priority-based scheduling scheme through mathematical analysis. Finally, the inability of the video server to completely observe the network state brings a challenge for video bitrate adjustments. 

\begin{figure}[tb]
\centering 
\includegraphics[height=2.1in,width=3.5in]{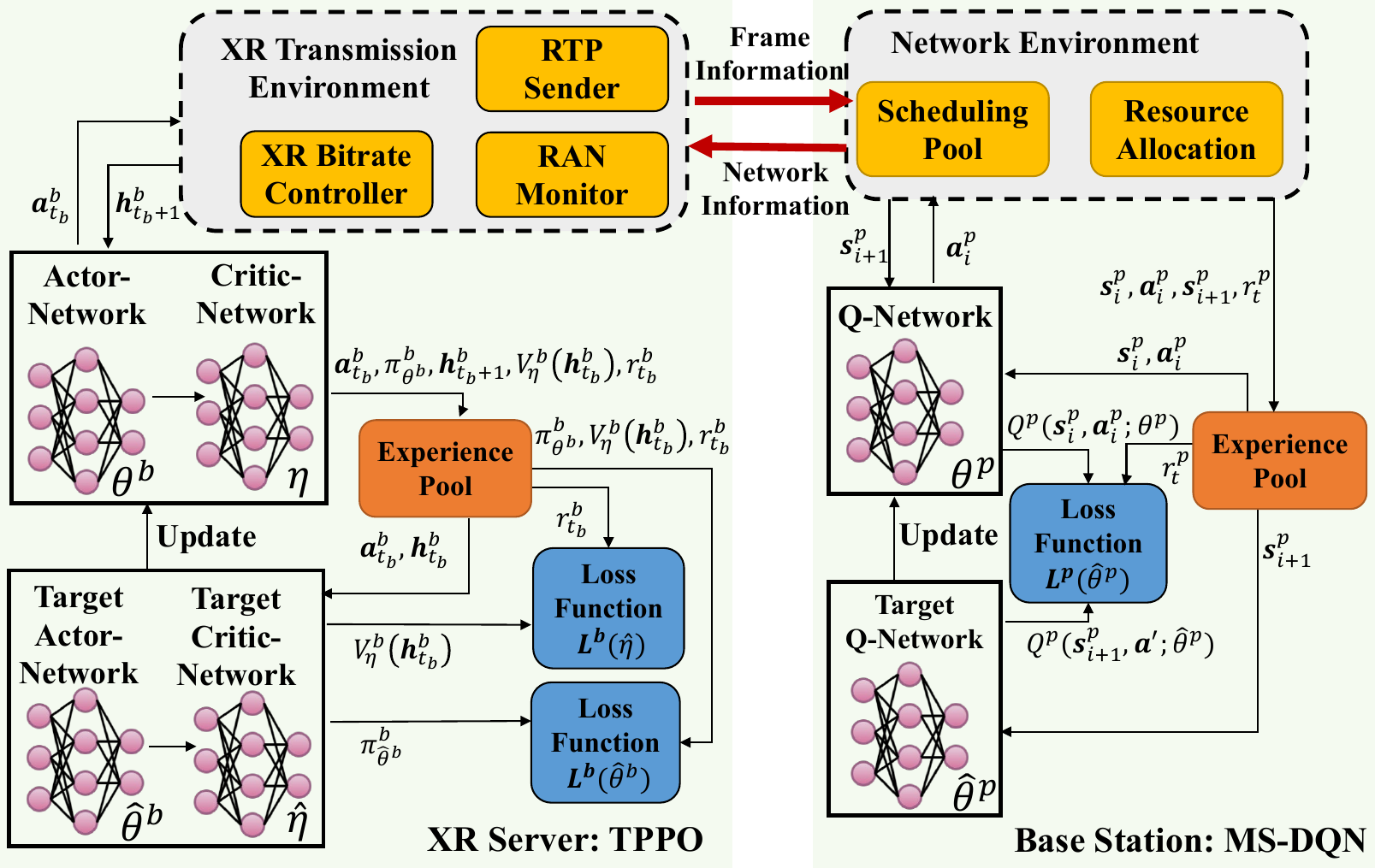}
\caption{Algorithmic diagram of the real-time XR transmission system.}
\label{fig:algorithm} 
\end{figure}

Therefore, we propose a multi-agent optimization solution that optimizes wireless resource scheduling and video bitrate allocation separately. The system diagram of the algorithm is shown in Fig. \ref{fig:algorithm}. The MS-DQN algorithm utilized on the BS side interacts with the wireless network environment for wireless resource allocation. The XR server employs the TPPO algorithm for the XR video bitrate adaptation. The BS and XR server exchange network and frame information with each other to improve video QoE. Specifically, the proposed MS-DQN is used to optimize the frame transmission success rate based on the frame importance and improve the transmission quality $\mathcal{Q}_{n,t_b}^t$. The TPPO algorithm is used to optimize video bitrate selection, thereby improving both video playback quality $\mathcal{Q}_{n,t_b}^p $ and transmission quality $\mathcal{Q}_{n,t_b}^t$.

\section{Frame-priority-based Scheduling Method in MAC Layer} \label{Sec:MSDQN}

In the MAC layer, the scheduler performs wireless resource scheduling for packets in the scheduling pool. Adopting an appropriate radio resource scheduling scheme can effectively enhance transmission quality. In this section, we describe the frame-priority-based scheduling scheme and formulate the problem to maximize the transmission quality. Then we introduce the proposed MS-DQN algorithm.

\subsection{Frame-Priority based Problem Formulation}

We first simplify the \textit{Problem} \ref{Prob:original} by representing it as a wireless resource allocation problem, and aim to improve the transmission quality.
\begin{Prob}
[Resource Allocation Subproblem] The wireless resource allocation subproblem can be represented as a scheduling problem based on frame priorities,
\label{prob:Resouce}
\begin{eqnarray} \mathop{\textrm{minimize}}_{{{\pi}^p}(\cdot)} && \! \! \! \! \sum_{n \in \mathcal{N} }\mathbb{E}[\mathcal{Q}_{n,t_b}^t]    \\
\mathop{\textrm{s.t.}} 
&& \! \! \! \! \eqref{equ:allocation},\eqref{equ:priority},\eqref{equ:qoe_t},\eqref{equ:RBconstraint},\nonumber
\end{eqnarray}
\end{Prob}

\begin{figure}[tb]
\centering 
\includegraphics[height=2.4in,width=3.4in]{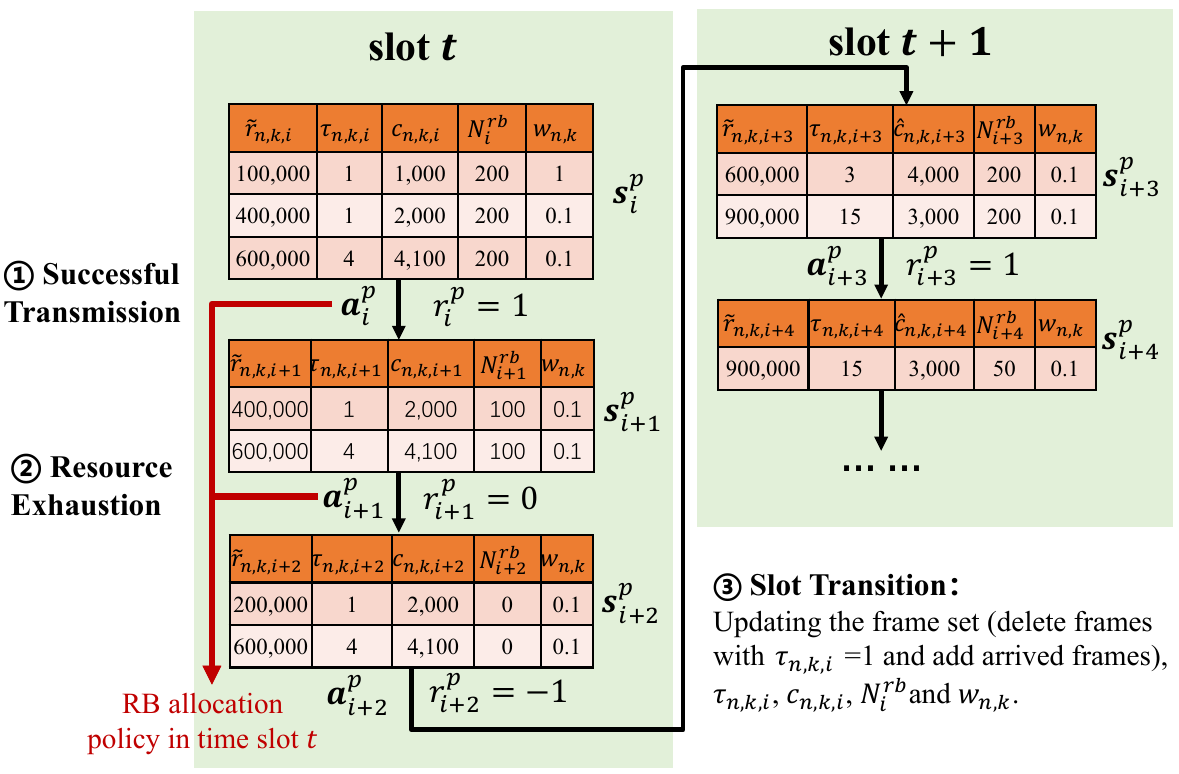}
\caption{This example demonstrates three types of state transitions. In slot $t$, the agent of MS-DQN makes two scheduling decisions and transmits two frames.}
\label{fig-priority} 
\end{figure}

We propose the MS-DQN based resource scheduling framework to solve the problem. When the action space is discrete and finite, DQN can work well. However, in the \textit{Problem} \ref{prob:Resouce}, the frame priority $p_{n,k,t}$ is a continuous value, and the action space is infinite. Although we can sort the priorities to select the scheduled users and discretize the action, the action space is still large. To reduce the action space, we proposed a multi-step based DQN algorithm. In our proposed MS-DQN algorithm, in each decision step, the agent outputs the probability values associated with each frame scheduling, which can be regarded as the priority of each frame. The proposed algorithm allows the agent to make multi-step decisions in each slot and select one frame of data to transmit in each decision step. Fig. \ref{fig-priority} shows an example of the state transition. In this example, the agent makes a three-step decision in slot $t$. Then the BS allocates RBs based on actions $\bm{a}^p_i$, $\bm{a}^p_{i+1}$, and $\bm{a}^p_{i+2}$. Then, the state transitions to $\bm{a}^p_{i+3}$, which is in slot $t+1$.

\subsection{MDP Reformulation of Problem $\mathbf{P}2$}

We define the \textit{Problem} \ref{prob:Resouce} as an infinite-horizon Markov decision process (MDP) by defining a tuple with five elements ${\mathcal{S}^p,\mathcal{A}^p,\mathcal{P}^p,\mathcal{R}^p,\gamma^p}$. Here, $\mathcal{S}^p$ is the state set, $\mathcal{A}^p$ is the action set, $\mathcal{P}^p$ is the transition probability, $\mathcal{R}^p$ is the reward function, and $\gamma^p$ is the discount factor $(0 < \gamma ^p< 1)$. The MDP is defined as follows,
\subsubsection{\textbf{State}} In each decision step $i$, we assume that the decision step is in time slot $t$, and $\mathcal{K}_i \subset \mathcal{K}_t$ represents the set of frames waiting for transmission. The state is represented by $\bm{s}_i^p=\{\bm{s}_{n,k,i}^p \ | \ \forall (n,k) \in \mathcal{K}_i\}$, where $\bm{s}_{n,k,i}^p=(w_{n,k,i},\widetilde{r}_{n,k,i},\tau_{n,k,i},{c}_{n,i},N^{rb}_i)$ consists of the data size waiting for transmission $\widetilde{r}_{n,k,i}=\widetilde{r}_{n,k,t} $, the RFDB $\tau_{n,k,i}=\tau_{n,k,t}$, the achievable data rate of each RB ${c}_{n,i}={c}_{n,t}$, the number of remaining RBs $N^{rb}_i$, and the frame importance weight $w_{n,k,i}=w_{n,k}$. Therefore, the state $\bm{s}_i^p$ can be represented as a state matrix of size $(|\mathcal{K}_i|,5)$.

\subsubsection{\textbf{Action}} In each decision step $i$, the agent selects one frame data to transmit. The action is given by $\bm{a}_i^p=(n_i,k_i) \in \mathcal{A}_i^p =\mathcal{K}_i$. Given an action, the number of RBs allocated to the selected frame is $\mathop{\textrm{min}} ( \lceil \frac{ \widetilde{r}_{n,k,i}}{c_{n,i}} \rceil,N_i^{rb})$.

\subsubsection{\textbf{Transition Probability}} Given the current state $\bm{s}_i$ and action $\bm{a}_i$, the transition probability $P(\bm{s}_{i+1}^p|\bm{s}_i^p,\bm{a}_i^p)$ can be defined. As depicted in Fig. \ref{fig-priority}, there are three types of state transitions.
\begin{itemize}
  \item{\textit{Type 1: Successful Transmission.}} With the chosen action, one frame is successfully transmitted regardless of whether RBs are used up.
  \item{\textit{Type 2: Resource Exhaustion}}. With the chosen action, RBs are used up, while the selected frame is not successfully transmitted.
  \item{\textit{Type 3: Slot Transition}}. When RBs are used up or the frame set is empty, the state transitions to a new state of the new slot by updating $w_{n,k,i}$, $\tau_{n,k,i}$, ${c}_{n,t}$, $N^{rb}_i$, and the frame set $K_i$ (drop frames with RFDB $\tau_{n,k,i}=1$ and add arrived frames). 
\end{itemize}
For type 1 and type 2, the transition probability $P(\bm{s}_{i+1}^p|\bm{s}_i^p,\bm{a}_i^p)$ is 1. For type 3, the transition probability $P(\bm{s}_{i+1}^p|\bm{s}_i^p,\bm{a}_i^p)$ depends on the frame arrival distribution, and the action does not affect the state transition.

\subsubsection{\textbf{Reward}} Under the selected action $\bm{a}^p_i=(n_i,k_i)$ at the decision step $i$, the reward function can be obtained based on the types of state transitions. Let $\widetilde{\mathcal{K}}_{n,t,i}$ denote the set of dropped frames
in decision step $i$. Therefore, we formulate the reward function as,
\begin{eqnarray}  
r^p(\bm{s}_i^p,\bm{a}_i^p)=\!  \left\{
{\begin{aligned}
0  \qquad \qquad \qquad \quad  , \  \mathop{\textrm{If type 1 or type 2}}, \\
   -\sum_{n \in \mathcal{N}} \! \sum_{k \in \widetilde{\mathcal{K}}_{n,t,i}} \!\!\!\! w_{n,k,i}, \qquad \qquad  \ \mathop{\textrm{If type 3}}. \\
% \\\mathop{\textrm{otherwise}} ,
\end{aligned}}  \right.
\end{eqnarray}

% \begin{eqnarray}
% \mathbb{I}(\lceil \frac{\widetilde{r}_{n_i,k_i,i}}{{c}_{n,i}} \rceil  \ge N^{rb}_i) \label{equ3-B-1}
% \end{eqnarray}

In this problem, the agent selects the highest priority frame for transmission. This policy determines the probability of the agent choosing each action at a given state, and can be expressed as $\pi^p(\bm{a}_i^p,\bm{s}_i^p)$. The criterion for selecting an action based on priority is similar to the policy used in MDP.

Given a policy $\pi$, the on-policy value function is defined as the expected total discounted reward,
\begin{eqnarray}\label{equ3-A-1}  V^p_{\pi}(\bm{s}^p)=\mathbb{E}_{\bm{a}_i^p\sim{\pi(\cdot|\bm{s}_i^p)} } [\sum_{i=0}^{+ \infty} (\gamma^p)^{i}r^p(\bm{s}_i^p,\bm{a}_i^p) |\bm{s}_0^p=\bm{s}^p].
\end{eqnarray}
The corresponding on-policy action-value function can be given by,
\begin{eqnarray}\label{on_policy_q} 
Q_{\pi}^p(\bm{s}^p, \! \bm{a}^p)
\! \! \! \! &=& \! \! \! \! \! \mathbb{E}^{\bm{a}_i^p\sim{\pi(\cdot|\bm{s}_i^p)}}_{\bm{s}_{i+1}^p\sim{P(\cdot|\bm{s}_i^p,\bm{a}_i^p)} } [\sum_{i=0}^{+ \infty} \! \! \gamma^{i}r(\bm{s}_i^p,\bm{a}_i^p) |\! \bm{s}^p_0=\bm{s}^p,\! \bm{a}^p_0=\bm{a}^p] \nonumber \\
\! \! \! \! &=&\! \! \! \! \! \mathbb{E}_{\bm{s}_{i+1}^p\sim{P(\cdot|\bm{s}_i^p,\bm{a}_i^p)} } [r(\bm{s}_i^p,\bm{a}_i^p) \nonumber\\ 
\! \! \! \! & & \! \! \! \! \! +\gamma\mathbb{E}_{{\bm{a}_{i+1}^p\sim{\pi(\cdot|\bm{s}_{i+1}^p)}}} [Q_{\pi}(\bm{s}_{i+1}^p,\bm{a}_{i+1}^p)]]. \!\!\!\!\!
\end{eqnarray}
According to \eqref{on_policy_q}, when the policy $\pi$ is different, the action-value function is also different. We define the optimal action-value function as $ Q_{\pi}^{p*}(\bm{s}^p,\bm{a}^p)=\mathop{\textrm{max}} _{\pi}  Q_{\pi}^p(\bm{s}^p,\bm{a}^p)$. Based on $ Q_{\pi}^{p*}(\bm{s}^p,\bm{a}^p)$, the optimal policy can be given by, 
\begin{eqnarray}  
\pi^{p*}(\bm{s}_i^p,\bm{a}_i^p)=\!  \left\{
{\begin{aligned}
1  \ , \  \mathop{\textrm{if}} Q_{\pi}^p(\bm{s}_i^p,\bm{a}_i^p)=\mathop{\textrm{max}}_{\bm{a}' \in \mathcal{A}_i^p}  Q_{\pi}^{p*}(\bm{s}^p_i,\bm{a}'),  \\
0 \ , \qquad \qquad   \ \mathop{\textrm{otherwise}} \qquad \qquad \qquad .
\end{aligned}}  \right.
\end{eqnarray}
Therefore, the Bellman equation for the optimal action-value function is, 
\begin{eqnarray}\label{Bellman_optimal_action_value}  Q_{\pi}^{p*}(\bm{s}^p_i,\bm{a}^p_i) = \mathbb{E}_{\bm{s}_{i+1}^p\sim{P(\cdot|\bm{s}_i^p,\bm{a}_i^p)} } [r(\bm{s}_i^p,\bm{a}_i^p) \nonumber \\ +\gamma \! \! \mathop{\textrm{max}}_{\bm{a}' \in \mathcal{A}_{i+1}^p} \! \! Q^{p*}(\bm{s}_{i+1}^p,\bm{a}')]. \nonumber
\end{eqnarray}

However, the optimal action-value function can not be obtained directly, because of the lack of prior information on transition probabilities $P(\bm{s}_{i+1}^p|\bm{s}_i^p,\bm{a}_i^p)$. A Q-learning method based on off-policy temporal difference is proposed to estimate the optimal action-value function $Q^*(s_i,a_i)$ iteratively by, 
\begin{eqnarray}\label{Qlearning}  Q^p(\bm{s}_i^p,\bm{a}_i^p) \! \! \! \! &=& \! \! \! \! Q^p(\bm{s}_i^p,\bm{a}_i^p)+\alpha[ r^p(\bm{s}_i^p,\bm{a}_i^p) \nonumber\\ 
\! \! \! \! & & \! \! \! \! + \gamma^p  \! \! \mathop{\textrm{max}}_{\bm{a}'\in \mathcal{A}_{i+1}^p}  \! \! Q(\bm{s}_{i+1}^p,\bm{a}')-Q^p(\bm{s}_i^p,\bm{a}_i^p)].
\end{eqnarray}

The neural network of DQN is parameterized by weights and biases denoted as $\theta^p$. To estimate the neural network, we can optimize the following loss functions $L^p(\theta_j^p)$ at iteration $j$,
\begin{eqnarray} 
L^p(\theta_j^p)=\mathbb{E}\bigg[\bigg(y(\hat{\theta}^p_j)-Q^p(\bm{s}_i^p,\bm{a}_i^p;\theta_j^p)\bigg)^2\bigg],
\end{eqnarray}
where
\begin{eqnarray}\label{equ:target}
y(\hat{\theta}^p)=r^p(\bm{s}_{i}^p,\bm{a}_{i}^p)+ \! \! \mathop{\textrm{max}}_{\bm{a}' \in \mathcal{A}_{i+1}^p}  \! \! Q^p(\bm{s}_{i+1}^p,\bm{a}';\hat{\theta}^p_j),
\end{eqnarray}
and $\hat{\theta}^p_j$ is the parameter of a fixed target network at each iteration $j$ \cite{cite:Dueling}. Therefore, the specific gradient update can be given by,
\begin{eqnarray} 
\nabla_{\theta_j^p}L(\theta_j^p)=\mathbb{E}\bigg[\bigg(y(\hat{\theta}^p_j)-Q(\bm{s}_i^p,\bm{a}_i^p;\theta_j^p)\bigg)\nabla _{\theta_j^p}Q(\bm{s}_i^p,\bm{a}_i^p;\theta_j^p)\bigg]. \nonumber 
\end{eqnarray}

\subsection{DQN Model Design and Training}

\begin{figure}[tb]
\centering 
\includegraphics[height=1.6in,width=3.4in]{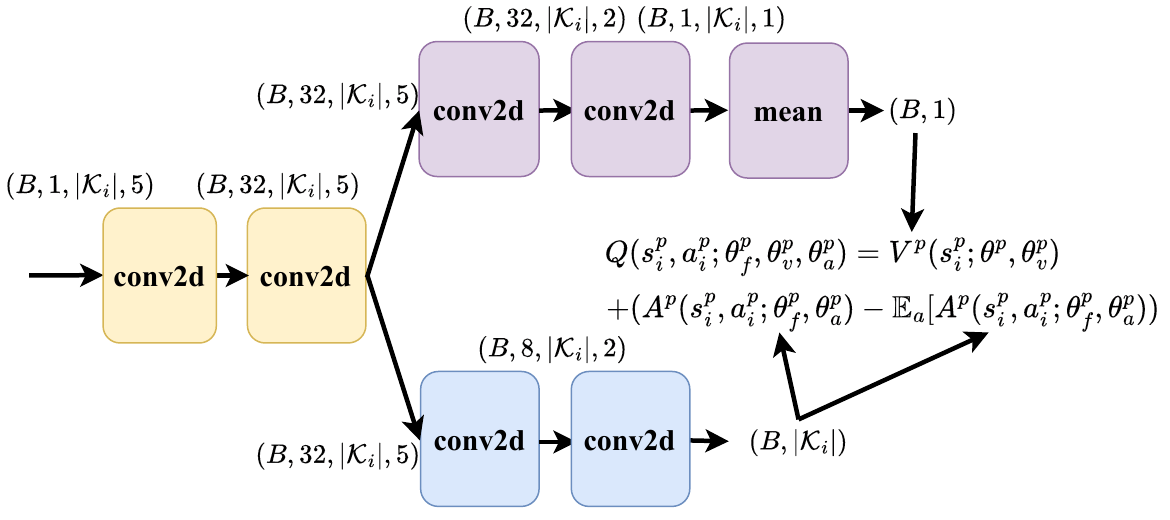}
\caption{2D-CNN-based neural network design. We estimate the value function and the advantage function respectively, then obtain the action value function. }
\label{fig:NN} 
\end{figure}

In our algorithm, two improvements are made to the original algorithm to develop an algorithm that is tailored to this task. Firstly, as previously mentioned, there are three types of state transitions $\pi^p(\bm{a}_i^p,\bm{s}_i^p)$. In types 1 and 2, rewards depend on both states and actions, while in type 3, they depend only on states. In order to better estimate the rewards brought by states and actions, we decouple the action-value function into a value function and an advantage function that quantifies the advantage of each action. More specifically, we have,
\begin{eqnarray} 
Q^p(\bm{s}_{i}^p,\bm{a}_{i}^p)=V^p(\bm{s}_{i}^p)+A^p(\bm{s}_{i}^p,\bm{a}_{i}^p),
\end{eqnarray}
where $A^p(\bm{s}_{i}^p,\bm{a}_{i}^p)$ denotes the advantage function. To implement this decoupling, we use the Dueling DQN method \cite{cite:Dueling}, and design the last module of the network as follows,
\begin{eqnarray} 
Q^p(\bm{s}_i^p,\bm{a}_i^p;\theta_f^p,\theta_v^p,\theta_a^p)  = V^p(\bm{s}_i^p;\theta_f^p,\theta_v^p)+ \qquad  \qquad \quad  \nonumber \\ 
\qquad \qquad \qquad \quad (A^p(\bm{s}_i^p,\bm{a}_i^p;\theta_f^p,\theta_a^p) 
-\mathbb{E}_a[A^p(\bm{s}_i^p,\bm{a}_i^p;\theta_f^p,\theta_a^p)
]), \nonumber
\end{eqnarray}
where $\theta_f^p$, $\theta_v^p$ and $\theta_a^p$ are parameters of feature extraction network, value network, and action network, respectively. 

Another challenge is that the input and output sizes of the network are not fixed. As mentioned earlier, the input size of the network is $(B,1,|\mathcal{K}_i|,5)$, and the output size is $(B,|\mathcal{K}_i|)$, where $B$ is the batch size. However, $|\mathcal{K}_i|$ can have different values in different states. The size of the action space and state space is dynamically changing. To overcome this, one solution is to use an upper bound $K$ for $|\mathcal{K}_i|$ (where $K \ge \mathcal{K}_i$) and pad the remaining input parameters with zeros \cite{cite:DDPG-padding}, or train different models for different input sizes directly \cite{cite:DDPGDNN}. These model design methods are not always general enough to be deployed in real systems.

Combining these two aspects, we designed our model based on 2D Convolutional Neural Network (2D-CNN) layers, as shown in Fig. \ref{fig:NN}. Two 2D-CNN layers extract features of the input state and output intermediate features with size $(B,32,|\mathcal{K}_i|,5)$. The value function network $\theta_v^p$ consists of two 2D-CNN layers and one mean-pooling layer, and the advantage function network $\theta_a^p$ consists of two 2D-CNN layers.

In each decision step, we store the agent's experience $(\bm{s}_i,\bm{a}_i,\bm{s}_{i+1},\bm{r}(a_i,s_i))$ in the replay buffer. When training the neural network, we sample data in batches from the replay buffer. We train the network using the $\epsilon$-greedy policy \cite{cite:Dueling}. Specifically, when the state transitions to type 3 and $\mathcal{K}_i=\emptyset$, we set the action to a default action $\bm{a}_i=(0,0)$, and let $Q(\bm{s}_i,\bm{a}_i)=0$.

\section{Learning-based XR Video Adaptation in application layer} \label{Sec:TPPO}

This section presents a learning-based adaptive bitrate algorithm for XR video applications. The adaptive bitrate algorithm performs bitrate selection at a large time scale in order to suit dynamic channel conditions. However, as a continuous action problem, we propose using a Transformer-based PPO algorithm to adjust video bitrate.

\subsection{Adaptive Bitrate Problem Formulation}

According to the \textit{Problem} \ref{Prob:original}, with a given resource allocation policy $\pi^p(\cdot)$, we can reformulate the problem as an adaptive bitrate problem for solving $\pi^b(\cdot)$,
\begin{Prob} 
[Adaptive Bitrate Subproblem] The adaptive bitrate subproblem can be represented as,
\label{prob:Bitrate}
\begin{eqnarray} \mathop{\textrm{maximize}}_{{{\pi}^b}(\cdot)} && \! \! \! \! \sum_{n \in \mathcal{N} }\mathbb{E}[\log \mathcal{Q}_{n,t_b}^p  + \mathcal{Q}_{n,t_b}^t],    \\
\mathop{\textrm{s.t.}} 
&& \! \! \! \!\eqref{equ:bitrate},\eqref{equ:qoe_p},\eqref{equ:qoe_t},\eqref{equ:Biterateconstraint} \nonumber
\end{eqnarray}
\end{Prob}

The adaptive bitrate problem aims to maximize the sum of the expected logarithm of the video quality and the transmission quality. We also consider the fairness of bitrate allocation to multiple users. 

To solve this problem, we propose a learning-based adaptive bitrate selection algorithm using the TPPO algorithm. The Transformer is used to extract historical semantic features related to the environment, enabling TPPO to choose the appropriate video bitrate utilizing the historical information. The TPPO algorithm is a type of stochastic policy gradient algorithm, which employs two modules, i.e., the actor network and the critic network. The actor network maximizes the objective function and uses the gradient descent algorithm to update the policy model. The critic network evaluates the objective function based on reward parameters, using the evaluation result as the reference for further performance improvement. By iteratively updating these networks, our algorithm can learn a good policy for bitrate selection in XR video applications.

\subsection{POMDP formulation of \textit{Problem} \ref{prob:Bitrate}}

Considering that the XR server cannot obtain all state information of the BS, We formulate the \textit{Problem} \ref{prob:Bitrate} as a POMDP. The more detailed POMDP definition of the adaptive bitrate problem is as follows,

\subsubsection{\textbf{Observed State}} In XR server, the state $\bm{s}^b_{t_b} \in \mathcal{S}^b$ is partially observable. Therefore, we define the observed state at bitrate adjustment period $t_b$ as $\bm{o}^b_{t_b} = (\hat{N}_{t_b}^{rb}, U_{n,t_b}^{rb}, C_{n,t_{b}}, |{\mathcal{K}}_{n,t_b}|, |\widetilde{\mathcal{K}}_{n,t_b}|, W_{n,t_b}, \widetilde{W}_{n,t_b}, n \in \mathcal{N}) \in \mathcal{O}^b$. The observed state $\bm{o}^b_{t_b}$ includes the average bandwidth $\hat{N}_{t_b}^{rb}$, RB utilization ${U}_{n,t_b}^{rb}$, average rate ${C}_{n,t_b}$, the number of transmitted frames $|{\mathcal{K}}_{n,t_b}|$, the number of failed frames $|\widetilde{\mathcal{K}}_{n,t_b}|$, the importance weights sum of the transmitted frames ${W}_{n,t_b}$ and the importance weights sum of the failed frames $\widetilde{W}_{n,t_b}$. These are calculated as follows,
\begin{eqnarray}
\hat{N}_{t_b}^{rb}&=&\sum_{t= t_bT_b}^{ t_bT_b+T_b-1} \frac{{N}_{t}^{rb}}{T_b}, \\
U_{n,t_b}^{rb} &=&{\sum_{t=t_bT_b}^{t_bT_b+T_b-1}   \sum_{(n,k) \in \mathcal{K}_{n,t}} \! \! \! \! \!  N_{n,k,t}^{rb}}, \\
C_{n,t_b} & = &  {\sum_{t=t_bT_b}^{t_bT_b+T_b-1}   \sum_{(n,k) \in \mathcal{K}_{n,t}} \! \! \! \! \!  c_{n,t}N_{n,k,t}^{rb}}, \\
W_{n,t_b} &=& \!  \sum_{(n,k) \in {\mathcal{K}}_{n,t_b}} \! \! \! \! \!  w_{n,k}, \\ 
\widetilde{W}_{n,t_b} &=& \!  \sum_{(n,k) \in \widetilde{\mathcal{K}}_{n,t_b}} \! \! \! \! \!  w_{n,k}.
\end{eqnarray}

\subsubsection{\textbf{Action}} Action is the adjustment policy of the video bitrate. At bitrate adjustment period $t_b$, the action is denoted by $\bm{a}^b_{t_b}=(\Delta R_{1,t_b} \cdots \Delta R_{N,t_b})$, where $\Delta R_{n,t_b} \in [-\Delta R^{max},\Delta R^{max}]$ is the adjustment value with limited range. Then the bitrate after adjustment is $R_{n,t_b}=R_{n,t_b-1}+\Delta R_{1,t_b}$.

\subsubsection{\textbf{Transition Probability}} With state $\bm{s}^b_{t_b}$ and action $\bm{a}^b_{t_b}$, the transition probability can be given by $P(\bm{s}^b_{t_b +1}|\bm{s}^b_{t_b}, \bm{a}^b_{t_b})$. POMDP also has partial observations $\bm{o}^b_{t_b}$, which are conditioned on the conditional observation probability $P(\bm{o}^b_{t_b}|\bm{s}^b_{t_b})$. Therefore, the transition distribution under the policy $\pi_{\theta^b}$ can be expressed as,
\begin{eqnarray} P(\Gamma|\pi^b)= P(\bm{s}^b_{0}) P(\bm{o}^b_{0}|\bm{s}^b_{0}) \!\!\!\!\!\!\! \prod_{t_b \in \mathcal{T}_b \verb|\| \{0\} } \!\!\!\!\!\!\!\! P(\bm{o}^b_{t_b}|\bm{s}^b_{t_b}) \quad \nonumber \\ 
\quad \prod_{t_b \in \mathcal{T}_b \verb|\| \{0\}} \!\!\!\!\!\!\!\! P(\bm{s}^b_{t_b}|\bm{s}^b_{t_b-1},\bm{a}^b_{t_b-1}) \!\!\!\!\!\! \prod_{t_b \in \mathcal{T}_b \verb|\| \{0\}} \!\!\!\!\!\!\!\! \pi_{\theta^b}(\bm{a}^b_{t_b-1}|\bm{h}^b_{t_b-1}),
\label{equ:trajectory} 
\end{eqnarray}
where $\Gamma=(\bm{s}^b_0,\bm{o}^b_0,\bm{a}^b_0,\cdots,\bm{s}^b_{|T_b|-1},\bm{o}^b_{|T_b|-1})$ is the trajectory, and $\bm{h}^b_{t_b}=(\bm{o}^b_0,\bm{a}^b_0,\cdots,\bm{o}^b_{t_b-1},\bm{a}^b_{t_b-1},\bm{o}^b_{t_b})$ is the historical sequence.

\subsubsection{Reward} Before the bitrate agent makes the $(t_b+1)^{th}$ decision, the XR bitrate controller can use the information from the XR server and the RAN monitor to calculate the reward. According to \textit{Problem} \ref{prob:Bitrate}, the reward function can be expressed as,
\begin{eqnarray} r^b_{t_b}= \sum_{n \in \mathcal{N}}(\log \mathcal{Q}_{n,t_b}^p  + \mathcal{Q}_{n,t_b}^t).\label{equ-BitrateReward}
\end{eqnarray}
$\mathcal{Q}_{n,t_b}^p $ can be obtained from the local state, and $\mathcal{Q}_{n,t_b}^t$ is calculated from the BS information collected in real-time.

The theory of policy gradient provides a way to optimize the parameters of a policy to maximize the expected return. In POMDP, the expected return is defined as the sum of rewards obtained by following a policy over a sequence of time steps. The update direction for the policy parameters is given by the gradient of the objective function. We derive the theory of policy gradient theory and ensure that the update direction is correct for the POMDP. The overall expected return is defined as,
\begin{eqnarray} J^b(\pi_{\theta^b})= \mathbb{E}_{\Gamma \sim{\pi_{\theta^b}}}[R(\Gamma)]=\sum_{\Gamma}P(\Gamma|\pi_{\theta^b})R(\Gamma).\label{equ-BitrateReturn}
\end{eqnarray}
where $R(\Gamma)=\sum_{t_b=0}^{\infty}\gamma^br^b_{t_b}$, $\gamma^b$ is the discount factor. In order to obtain a policy $\pi_{\theta^b}$ that maximizes the expected return $J^b(\pi_{\theta^b})$, we update the parameters $\theta^b$ using the gradient of the objective function. The correct update direction is given by,

\textit{Theorem 1}: Given a function approximator $Q^b_{\eta}$, if the following conditions are satisfied, \\
\qquad \ \ \  1) $Q^b_{\eta}(\bm{h}^b_{t_b},\bm{a}^b_{t_b})=\nabla_{\theta^b} \log \pi^b_{\theta^b}(\bm{a}^b_{t_b}|\bm{h}^b_{t_b})^\mathsf{T}w,$\\
\qquad 2) the parameters $w$ are obtained by minimizing the mean-squared error, $\mathbb{E}_{\Gamma \sim{\pi_{\theta^b}^b}}[(Q^b_{\eta}(\bm{h}^b_{t_b},\bm{a}^b_{t_b})-Q^b_{\theta^b}(\bm{h}^b_{t_b},\bm{a}^b_{t_b}))^2]$,\\
the policy gradient of POMDP is given by,
\begin{align}
\nabla_{\theta^b} J^b(\pi_{\theta^b}^b)=\mathbb{E}_{\Gamma \sim{\pi_{\theta^b}^b}}[\nabla_{\theta^b} \log \pi_{\theta^b}^b(\bm{a}^b_{t_b}|\bm{h}^b_{t_b})Q^b_{\eta}(\bm{h}^b_{t_b},\bm{a}^b_{t_b})]. \label{equ-POMDPpolicygradient} \nonumber
\end{align}

\textit{Proof:} Please refer to Appendix A.

Therefore, the actor-critic architecture can also be extended to the partial observations domain.

\subsection{Model Design and Training}

TPPO is a variant of the PPO \cite{cite:PPOoriginal} algorithm. For both the actor network and the critic network, we use a 6-layer Transformer encoder to build feature extraction and then use a fully connected layer for outputting results. The input size of the Transformer encoder is $7N$, and the head of self-attention is set to the number users $N$. The size of output layer of the actor network and the critic network are $N$ and 1, respectively. We sequentially input the information in the observation window into the actor network and critic network to obtain features at different time scales.

Similar to PPO \cite{cite:PPOoriginal}, TPPO uses a clipped objective function to update the policy parameters, which ensures stable policy improvement while preventing large policy updates. TPPO utilizes the Transformer model to extract semantic information from the historical state, including network status and bitrate control information. By leveraging this information, we can perform video bitrate control and estimate the state-value function, which is critical for achieving optimal performance in adaptive real-time XR bitrate streaming. With the use of the Transformer model, our TPPO algorithm is able to capture complex temporal dependencies in the input data and achieve superior performance compared to existing bitrate adaptation algorithms.

% After obtaining the hidden feature vectors, TPPO applies a policy network to predict the distribution of the next bitrate level. The policy network is also implemented using transformer architecture, which takes the hidden feature vectors as input and outputs the action probabilities.

The TPPO algorithm also uses the importance sampling method \cite{cite:PPOoriginal}, and obtains the training data through old actor ($\hat{\theta}^b$) and critic network ($\hat{\eta}$). Therefore, the loss of the actor network is, 
\begin{align}
L^b(\theta^b)=-\mathbb{E}[min(\frac{\pi^b_{\theta^b}}{\pi^b_{\hat{\theta}^b}}A_{t^b}(\bm{h}^b_{t_b},\bm{a}^b_{t_b}), \nonumber \\ 
clip(\frac{\pi^b_{\theta^b}}{\pi^b_{\hat{\theta}^b}},1-\epsilon,1+\epsilon)A_{t^b}(\bm{h}^b_{t_b},\bm{a}^b_{t_b})] 
\end{align}
where $\epsilon$ is a hyperparameter, and $clip(\cdot)$ is the clip function. The clip function makes the two distributions not much different, and guarantees the update rate is under control. $A_{t^b}(\bm{h}^b_{t_b},\bm{a}^b_{t_b})$ is the generalized advantage function, and can be expressed as,
\begin{align} 
& \quad A_{t^b}(\bm{h}^b_{t_b},\bm{a}^b_{t_b})= Q^b(\bm{h}^b_{t_b},\bm{a}^b_{t_b}) - V^b_{\eta}(\bm{h}^b_{t_b}) \\ 
\text{where} & \quad Q^b(\bm{h}^b_{t_b},\bm{a}^b_{t_b})=\sum_{t_b'=t_b}^{+\infty }(\gamma^b\phi)^{t_b'-t_b}\delta_{t_b'} + V^b_{\eta}(\bm{h}^b_{t_b}),  \nonumber \\
& \quad
\delta_{t_b}=r^b_{t_b}+\gamma_b V^b_{\eta}(\bm{h}^b_{t_b+1})-V^b_{\eta}(\bm{h}^b_{t_b}), \quad \nonumber
\end{align}
$\phi$ is used to adjust the bias-variance tradeoff, and $V^b_{\eta}(\bm{h}^b_{t_b})$ is the state-value function estimated by the critic network. Therefore, we train the critic network by minimizing the mean squared error of the state value function,
\begin{align}
L^b(\hat{\eta})=\mathbb{E}[\sum_{\tau=0}^{+\infty} \gamma_b^{\tau} r^b_{t_b+\tau}-V^b_{\hat{\eta}}(\bm{h}^b_{t_b})]. \label{equ-Losscritic}
\end{align}
% \begin{align}
% L^b(\eta)=\mathbb{E}[r^b_{t_b}+\gamma_b V^b_{\eta}(\bm{h}^b_{t_b+1})-V^b_{\eta}(\bm{h}^b_{t_b})]. \label{equ-Losscritic}
% \end{align}

\section{EXPERIMENTAL RESULTS } \label{Sec:Result}

    % In this section, we evaluate the performance of the proposed scheme. For all results, unless specified otherwise, the bandwidth is 100MHz with sub-carrier spacing (SCS) 30KHz. We consider the Urban Microcell deployment scenario. users distributed in the area within [500m, 500m] randomly. In our system, the bitrate of the XR video is assumed to be 30Mbps, and the frame rate is 120FPS. Therefore, the mean size of each frame is 250Kbits. The jitter of frame size and arrival time is generated according to \cite{cite:TS38838}. The FDB is set to be 10ms. Results are obtained through a 2000-slot simulation.

    In this section, we evaluate the performance of the proposed algorithm. For all results, unless specified otherwise, the bandwidth of each RB is 360KHz, the transmission power of each RB is 10 mW. We consider the pathloss of the Urban Micro cell deployment scenario \cite{cite:TS38901}, and the noise power spectral density is -174 dBm/Hz. Users are distributed in the area within [500m, 500m] randomly. For the XR video, the bitrate of the XR video is assumed to be between 1Mbps and 30Mbps, and the frame rate is 60FPS. Therefore, the mean size of each frame is 250Kbits. The jitter of frame size and arrival time is generated according to \cite{cite:TS38838}. I-frames are transmitted periodically every $K=4$ frame, while the other frames are P-frames. The average size ratio between one I-frame and one P-frame is $\alpha=1.5$ \cite{cite:TS38838} and the importance weights of I-frames and P-frames are set to 1 and 0.1, respectively. The FDB is set to be 10ms. The distribution of the first frame's arrival time can be represented as $t_{n,0} \in [0,1000/f_n]$ ms. The length of one slot $\Delta t$ is 1 ms, and the time interval of bitrate adjustment period is 100 ms. For the QoE model, we set $\alpha_p=0.2$, $\alpha_t=0.8$, $N_t=150$, $\beta_1=60$, $\beta_2=\frac{1}{2}$. For the proposed MS-DQN algorithm, the discount factor $\gamma^p =0.9$, and the learning rate is $0.001$. For the proposed TPPO algorithm, the discount factor $\gamma^b =0.9$, $\epsilon =0.2$, $\phi=1$, and the learning rate of the actor network and the critic network is $0.0002$ and $0.0001$, respectively.

% Results are obtained through a 2000-slot simulation.

\begin{figure}[tb]
\centering 
\includegraphics[height=2.0in,width=2.5in]{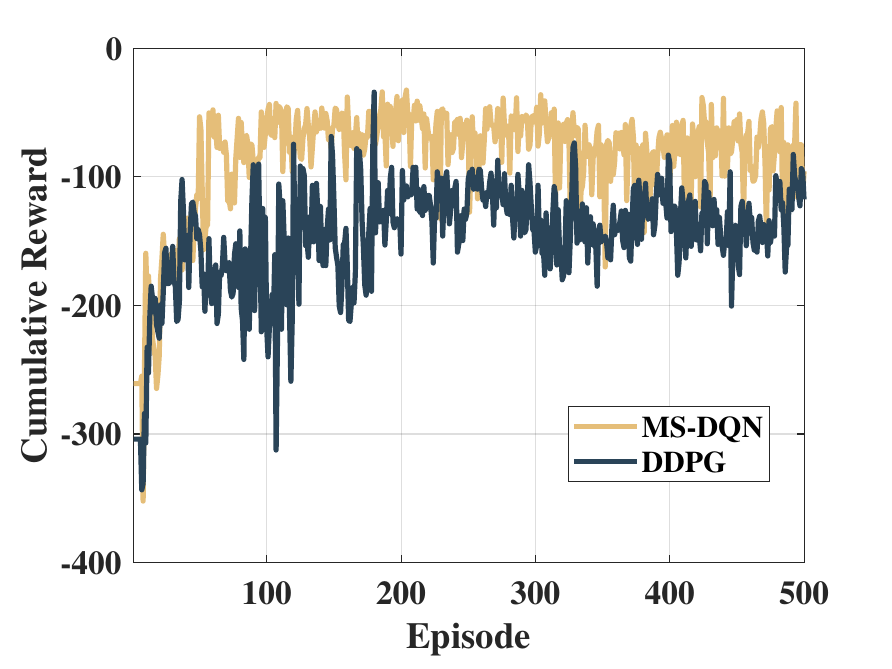}
\caption{Convergence performance of MS-DQN and DDPG algorithms for wireless resource allocation problem.}
\label{fig:TransConv} 
\end{figure}

\subsection{Performance Comparison for BS Scheduling} 
To compare BS scheduling algorithms, we fix the bitrate of XR video to 20Mbps, and compare the proposed MS-DQN algorithm with baseline scheduling algorithms. Results are obtained through a 2000-slot simulation. We consider the following baseline algorithms. 
\subsubsection{\textbf{DDPG\cite{cite:priority-DDPG}}} The DDPG algorithm scheduler uses a neural network to adjust the off-policy deterministic policy gradient algorithm based on an Actor-Critic framework. 
\subsubsection{\textbf{PF\cite{PF}}} The PF scheduler considers the user's current rate and historical average rate to calculate the priority.
\subsubsection{\textbf{PF-I\cite{cite:frameintegrated}}} The Frame-level integrated based PF scheduler is designed to take into account both the integrity of the frame and the network channel conditions for users.

We first compare the convergence performance between our proposed MS-DQN algorithm and the DDPG algorithm However, when the number of users and the initial time of the first frame are different, the convergence curve will also be different. For the fairness of the comparison, in each episode, we train the neural network using scenarios with varying numbers of users and initial times and test with the same scenario to observe changes of cumulative reward. Each episode of training and testing is 2000 slots. Fig. \ref{fig:TransConv} depicts the cumulative reward of different algorithms. The number of users is 8. Both the proposed MS-DQN-based algorithm and the DDPG algorithm can converge after about 50 episodes. However, compared with the DDPG algorithm, the MS-DQN algorithm's cumulative reward is higher, which shows the proposed MS-DQN algorithm performs better than DDPG.

\begin{figure}[tb]
\centering 
\includegraphics[height=2.0in,width=2.5in]{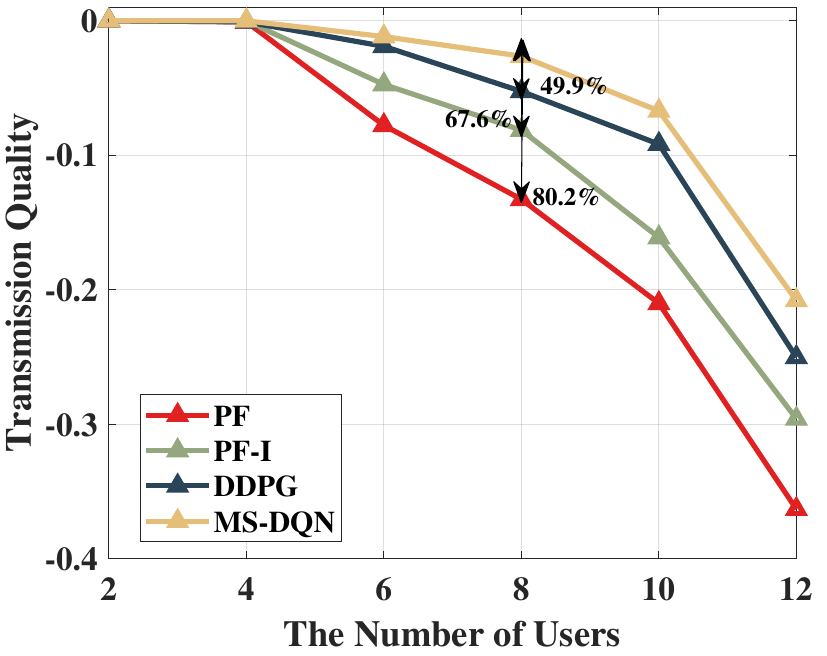}
\caption{Performance comparison of transmission quality between the proposed MS-DQN algorithm and baseline algorithms.}
\label{fig:TransQualityResultUE} 
% \vspace{-5mm}
\end{figure}

Fig. \ref{fig:TransQualityResultUE} shows the comparison of transmission quality among different algorithms under different numbers of users. We repeat the experiment 50 times, and each experiment is performed for 2000 time slots. As shown in Fig. \ref{fig:TransQualityResultUE}, the proposed MS-DQN algorithm outperforms the baseline algorithm. For example, when the number of users is 8, compared with PF, PF-I, and DDPG algorithms, the proposed MS-DQN has an improvement of 80.2\%, 67.6\%, and 49.9\% in transmission quality, respectively.

\begin{figure}[tb]
\centering 
\includegraphics[height=2.0in,width=2.5in]{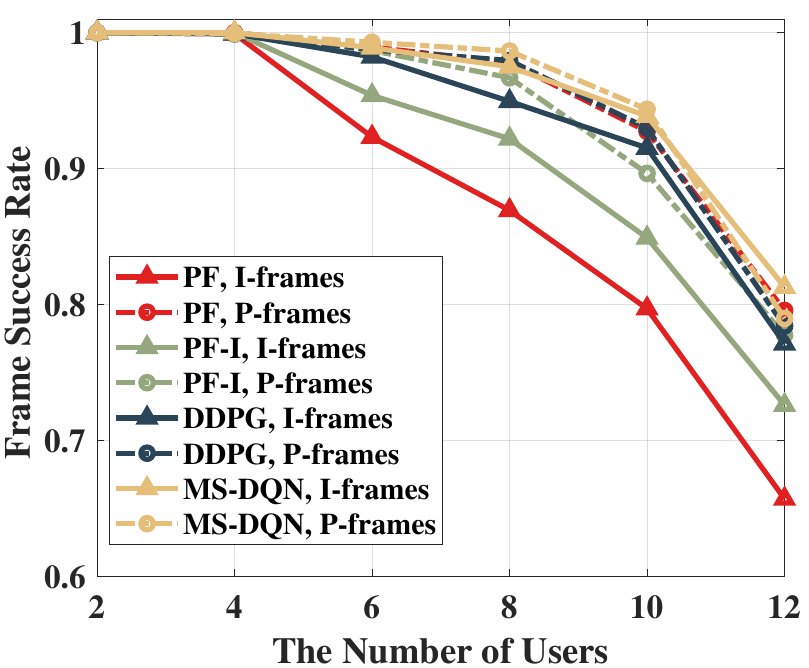} 
\caption{I-frames and P-frames transmission performance comparison between the proposed MS-DQN algorithm and baseline algorithms.}
\label{fig:TransQualityResultIPFrame}
\vspace{-3mm}
\end{figure}

 Fig. \ref{fig:TransQualityResultIPFrame} illustrates the impact of different algorithms on the frame success rates of I-frames and P-frames. Specifically, for I-frame transmission, compared to the PF, PF-I, and DDPG algorithms, the proposed MS-DQN algorithm can improve the frame success rate by 10.6\%, 5.3\%, and 2.5\% respectively when the number of users is 8. For P-frame transmission, MS-DQN achieves a frame success rate improvement of 0.7-2.0\% when the number of users is 8.

\begin{figure}[tb]
\centering 
\includegraphics[height=2.0in,width=2.5in]{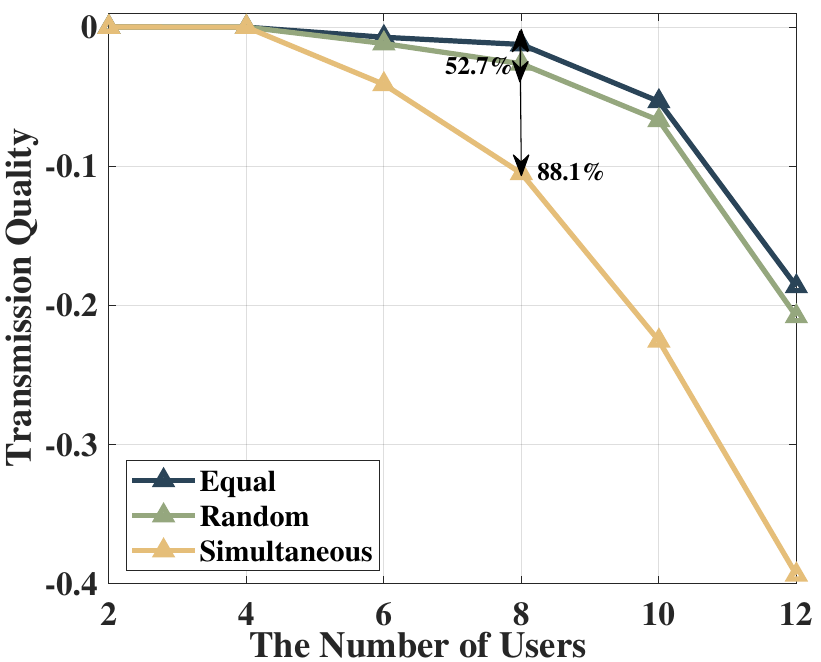}
\caption{Performance comparison of transmission quality with the different initial arrival time. It indicates that the distribution of frame arrivals will affect the transmission quality.}
\label{fig:TransQualityResultArrivalTime} 
\vspace{-3mm}
\end{figure}

Considering the impact of the initial arrival time, we test three different scenarios and plot Fig. \ref{fig:TransQualityResultArrivalTime}. We consider three scenarios. Scenario 1: the initial frame arrival follows a uniform distribution. (Random); Scenario 2: the initial frames of all users arrive simultaneously (Simultaneous); Scenario 3: the initial frames of all users arrive with equal intervals (Equal). As shown in Fig. \ref{fig:TransQualityResultArrivalTime}, when the number of users is 8, compared with the Random and Simultaneous settings, the Equal setting can achieve an 88.1\% and 52.7\% increase in transmission quality, respectively. This indicates that the frame arrival interval of different users can affect the frame success rate of BS transmission. XR video servers can use this feature to arrange and adjust encoders to optimize real-time XR video transmission.

\subsection{Performance Comparison for Bitrate Adaption}

\begin{figure}[tb]
\centering 
\includegraphics[height=2.0in,width=2.5in]{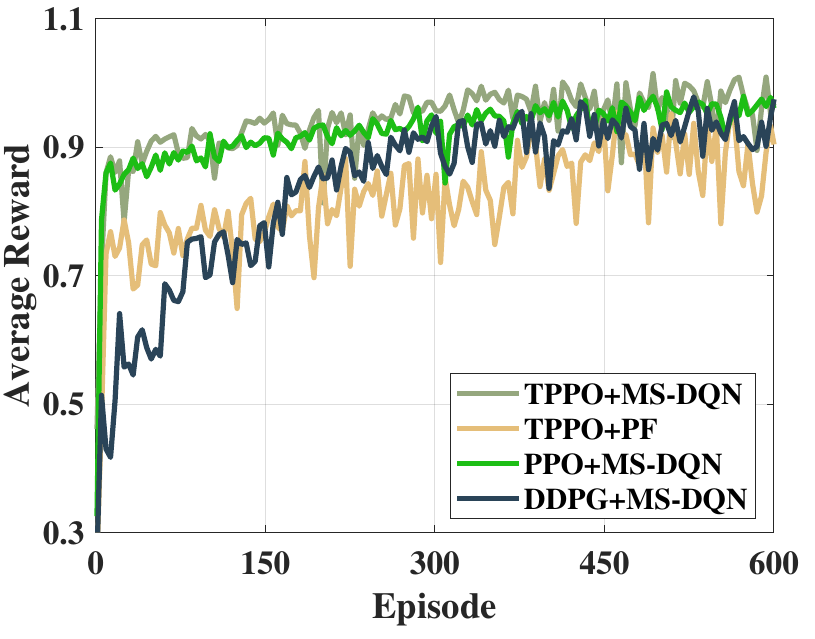}
\caption{Convergence performance of TPPO+MS-DQN and baseline algorithms.}
\label{fig:ABRCov} 
\vspace{-3mm}
\end{figure}

Then we compare the TPPO algorithm with the baseline algorithm and obtain the QoE optimization results. The experimental results were obtained through simulation over 1000 bitrate adjustment periods, with each bitrate adjustment period consisting of 100 slots. We consider following three baseline algorithms,
\subsubsection{\textbf{PPO \cite{cite:PPOABR}}} The PPO algorithm uses a fully connected network instead of the Transformer architecture.
\color{black}
\subsubsection{\textbf{DDPG \cite{cite:DDPGABR}}} Unlike the PPO algorithm, DDPG uses deterministic policy gradient to obtain the optimal actions.
\subsubsection{\textbf{Search}} The Search algorithm assumes that the XR bitrate controller has all prior information before the next bitrate adjustment period, including the maximum number of RBs and the channel state information. It uses the binary search algorithm to select the bitrate that maximizes the current QoE. It is important to note that the Search algorithm can only provide an approximation of optimal performance as a reference. The unavailability of prior information in practical systems makes it unusable.

Fig. \ref{fig:ABRCov} shows the convergence of algorithms with Episodes. We train the networks of TPPO using randomly generated channels and XR video parameters, then test it using fixed channels and XR video parameters to ensure fairness in comparison. We execute a test after every 4 training episodes. Average QoE represents the mean QoE for each user during each bitrate adjustment period, with TPPO+MS-DQN delivering the highest QoE performance. Fig. \ref{fig:ABRCov} shows that TPPO+MS-DQN, PPO+MS-DQN and TPPO+PF algorithms converge after around 200 episodes, while DDPG+MS-DQN reaches convergence after around 300 episodes. This implies that TPPO is capable of achieving better convergence. Although the convergence speed of PPO is similar to that of TPPO, the performance of the PPO algorithm is slightly worse than that of TPPO.

\begin{figure}[tb]
\centering 
\includegraphics[height=2.0in,width=2.5in]{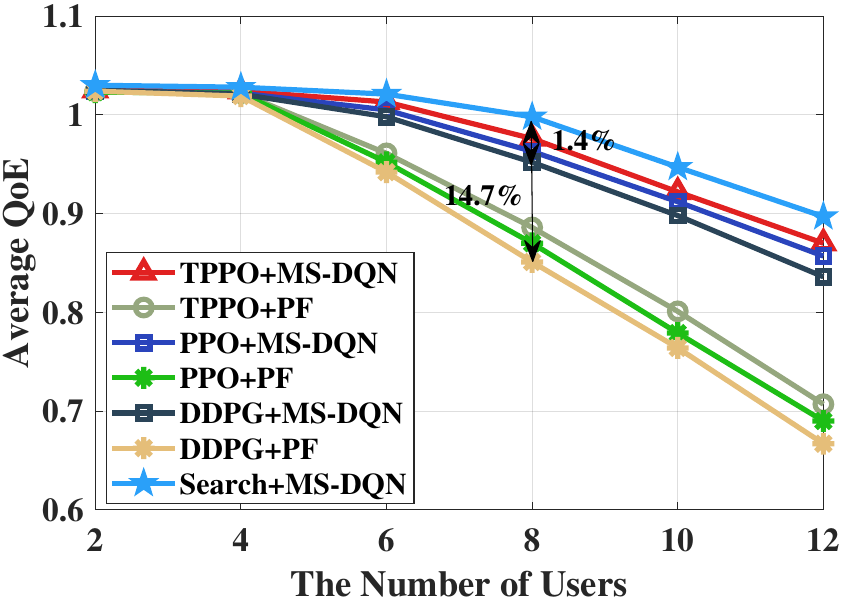}
\caption{The comparison of average QoE results under the different number of users.}
\label{fig:QOECompareUE} 
\vspace{-3mm}
\end{figure}

Fig. \ref{fig:QOECompareUE} shows the comparison of the average QoE under different numbers of users. As shown in the figure, the proposed TPPO+MS-DQN algorithm performs the best compared to the baseline algorithm. When the number of users is less than or equal to 4, the average QoE of users is similar because the BS can satisfy real-time XR video transmission with the maximum bitrate. With the increase in user number, competition for resources ultimately results in a decrease in the mean QoE. When the number of users is 8, the proposed TPPO+MS-DQN algorithm improves the average QoE by 1.4\%-14.7\% compared to the baseline algorithms. Specifically, the TPPO+MS-DQN algorithm enhances the average QoE by 14.7\% compared to the DDPG+PF algorithm. Compared with the PPO+MS-DQN algorithm, the proposed algorithm improved the average QoE by 1.4\% using the transformer architecture. The search algorithm outperforms the TPPO algorithm as it can choose the optimal video bitrate based on prior knowledge of the network environment. The search algorithm can approach the optimal solution, but it cannot be used in practical systems because the practical systems are unable to obtain network information about the present and future. When the number of users is 8, the TPPO algorithm can achieve a 97.8\% average QoE of the Search algorithm.

\begin{figure}[tb]
\centering 
\includegraphics[height=2.0in,width=2.5in]{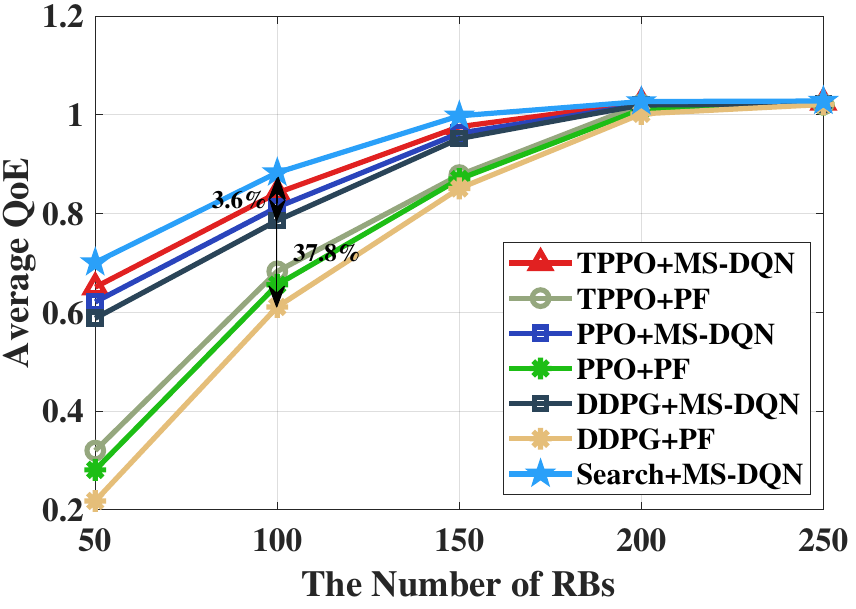}
\caption{The comparison of average QoE results under the different number of RBs.}
\label{fig:QOECompareBW} 
\vspace{-3mm}
\end{figure}

Fig. \ref{fig:QOECompareBW} shows the variation of average QoE under the different number of RBs (different bandwidths). The number of users is 8. As the figure shows, a larger bandwidth brings gains in the average QoE. When the number of RBs is greater than 200, the QoE of users remains constant, as the bandwidth can meet the transmission requirements of users' maximum bitrate. When the number of RBs is 100, the proposed TPPO+MS-DQN algorithm can achieve a performance improvement of 3.6\%-37.8\% and an 95.4\% average QoE of the Search algorithm.

\begin{figure}[tb]
\centering 
\includegraphics[height=3.0in,width=3.3in]{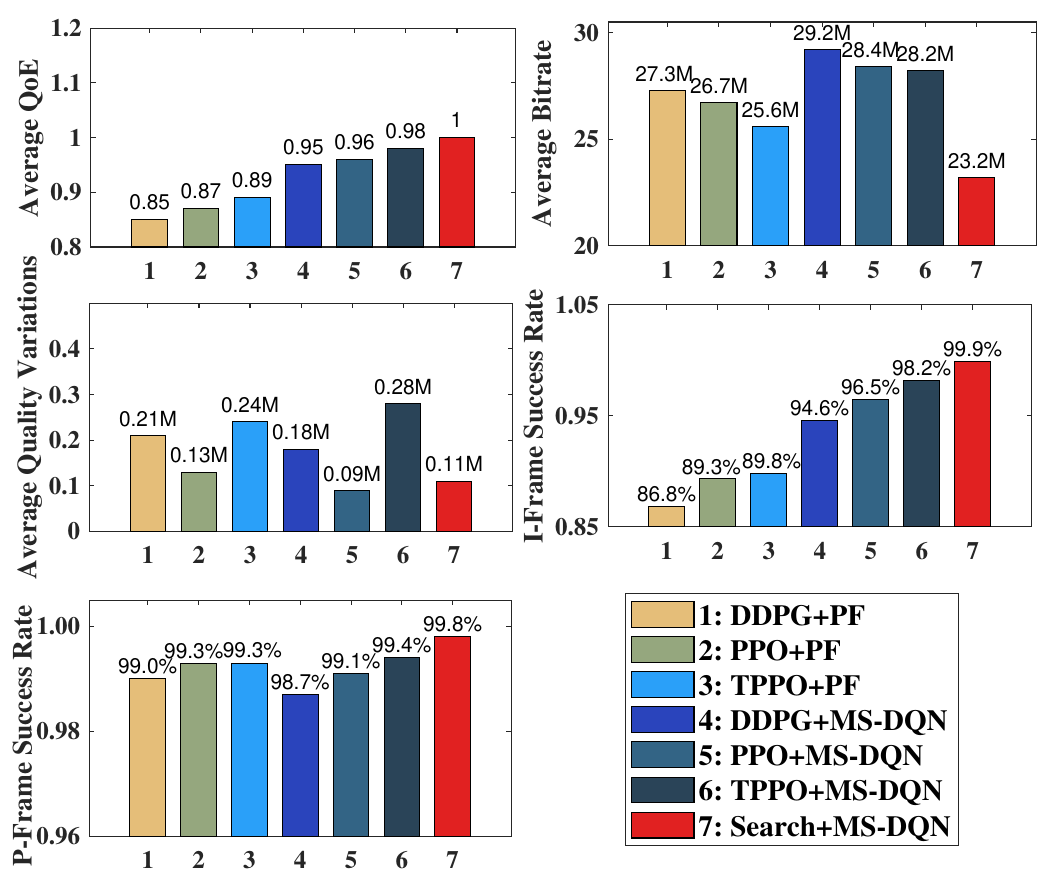}
\caption{Comparison of the various components of QoE when the number of users is 8 and the number of RBs is 150.}
\label{fig:QoEResultDetail} 
\vspace{-3mm}
\end{figure}

In Fig. \ref{fig:QoEResultDetail}, we plot the average QoE and the components that makeup QoE, including average bitrate, average quality variation, and frame transmission success rate. The proposed TPPO+MS-DQN algorithm can achieve an average bitrate of 28.2 Mbps, an average quality variation of 0.28 Mbps, and the success rates of transmitting I-frames and P-frames are 98.2\% and 99.4\%, respectively. The I-frame success rate of the DPPG+PF, PPO+PF and TPPO+PF are 86.8\%, 89.3\% and 89.8\%, which is lower than those for algorithms based on MS-DQN, because the PF algorithm doesn't consider the frame-priority. Compared with the DDPG+MS-DQN algorithm, the proposed TPPO+MS-DQN algorithm has a decrease in the average bitrate of 1.0 Mbps, but the success rates of transmitting I-frames and P-frames are increased by 3.6\% and 0.7\%, respectively. It is acceptable because frame transmission failures often result in greater user experience loss. Compared with the PPO+MS-DQN algorithm, the proposed TPPO+MS-DQN algorithm increases the transmission success rate of I-frames and P-frames by 1.7\% and 0.3\%, respectively, under the condition of almost constant average bitrate. It indicates the effectiveness of the transformer architecture.

\begin{figure}[tb]
\centering 
\includegraphics[height=2.0in,width=2.5in]{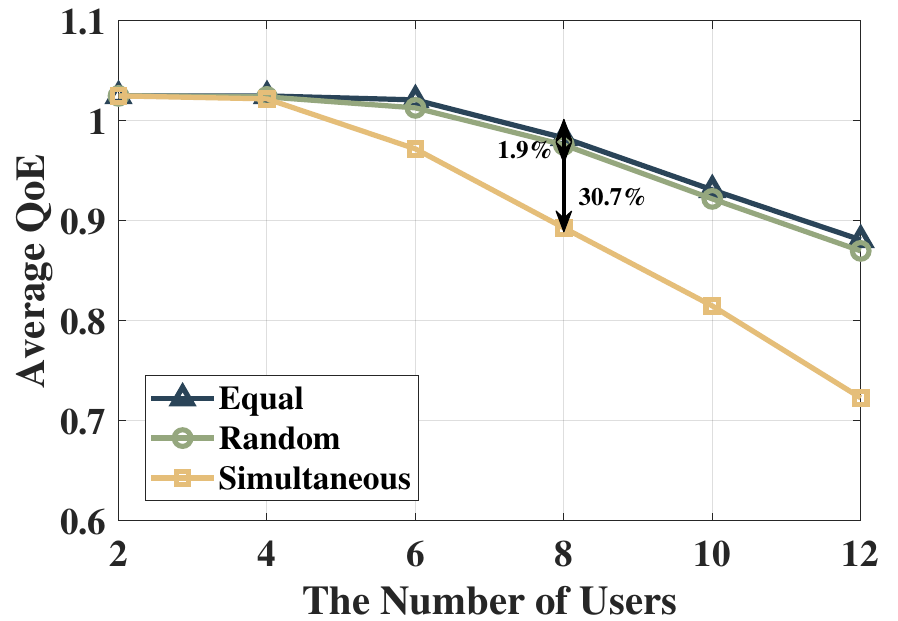}
\caption{The average QoE comparison results under the different initial arrival time.}
\label{fig:QualityResultArrivalTime} 
\vspace{-3mm}
\end{figure}

Considering the impact of the initial arrival time, Fig. \ref{fig:QualityResultArrivalTime} shows the average QoE comparison results under the different initial arrival time. We conduct tests in three scenarios, i.e., Equal, Random, and Simultaneous. As shown in Fig. \ref{fig:QualityResultArrivalTime}, the arrival time of the initial frame will affect the QoE of XR video transmission. When the number of users is 8, compared with the Random and Simultaneous settings, the Equal setting can improve the QoE by 1.9\% and 30.7\%, respectively. Therefore, XR servers can optimize QoE among multiple users by adjusting the initial arrival time. It is equivalent to adjusting the average transmission interval between video frames of different users. It provides a new approach for optimizing the QoE in multi-user XR transmission. We will expand the related work in the future.

\section{Conclusion} \label{Sec:Conclusion}
In this paper, we consider cross-layer transmission optimization for real-time XR video between XR servers and BS. We design a cross-layer transmission framework that supports information interaction between BSs and XR servers to assist them in wireless resource scheduling and adaptive bitrate adjustment. We propose a frame-priority-based scheduling method and design an MS-DQN-based scheduling algorithm for the BS. We also propose a TPPO-based adaptive bitrate algorithm for the XR server, which uses the semantic information of the historical environment to select an appropriate video bitrate while considering user fairness. This multi-agent-based method solves the problem of mismatched time scales. The experiments show that our TPPO+MS-DQN algorithm can effectively improve QoE for XR video transmission in multi-user scenarios. 

In addition, we demonstrate through experiments that the initial arrival time distribution of video frames affects the QoE of multi-user real-time XR video transmission. Therefore, we can optimize video transmission QoE by adjusting the initial frame arrival distribution. This research topic remains to be addressed in future work.


\bibliographystyle{IEEEtran}

\bibliography{reference}

\end{document}

\title{Quality of Experience Oriented Cross-layer Optimization for Real-time XR Video Transmission}


\author{Guangjin Pan, Shugong Xu, \IEEEmembership{Fellow, IEEE}, Shunqing Zhang, \IEEEmembership{Senior Member, IEEE},\\ Xiaojing Chen, \IEEEmembership{Member, IEEE} , and Yanzan Sun, \IEEEmembership{Member, IEEE} \\

}

\markboth{IEEE TRANSACTIONS ON CIRCUITS AND SYSTEMS FOR VIDEO TECHNOLOGY, ~Vol.~XX, No.~XX, June~2023}%
{Shell \MakeLowercase{\textit{et al.}}: Bare Demo of IEEEtran.cls for IEEE Transactions on Magnetics Journals}

\maketitle

\IEEEdisplaynontitleabstractindextext

\IEEEpeerreviewmaketitle

\appendices
\section{Proof of Theorem 1}
With full observations, the policy gradient theorem is proved in \cite{cite:PPOtheory}.

\textit{Lemma 1}: Given a function approximator $Q^b_{\eta}$, the policy gradient of MDP is given by,
\begin{align}
\nabla_{\theta} J^{b\star}(\pi_{\theta})=\mathbb{E}_{\Gamma^\star \sim{\pi_{\theta}}}[\nabla_{\theta} \log \pi_{\theta}(\bm{a}^b_{t_b}|\bm{s}^b_{t_b})Q^b_{\eta}(\bm{s}^b_{t_b},\bm{a}^b_{t_b})]. \label{equ-MDPpolicygradient} \nonumber
\end{align}
if the following conditions are satisfied, \\
1) $Q^b_{\eta}(\bm{s}^b_{t_b},\bm{a}^b_{t_b})=\nabla_{\theta} \log \pi_{\theta}(\bm{a}^b_{t_b}|\bm{s}^b_{t_b})^\mathsf{T}w,$\\
\qquad 2) the parameters $w$ are obtained by minimizing the mean-squared error, $\mathbb{E}_{\Gamma^\star \sim{\pi_{\theta}}}[(Q^b_{\eta}(\bm{s}^b_{t_b},\bm{a}^b_{t_b})-Q^b_{\pi_\theta}(\bm{s}^b_{t_b},\bm{a}^b_{t_b}))^2]$,\\

We compare the police gradient for MDP and POMDP as follows,
\begin{align}
\nabla_{\theta} J^{b\star}(\pi_{\theta})=\nabla_{\theta}  \sum_{\Gamma^\star} P[\Gamma^\star|\pi_{\theta} ]R(\Gamma^\star). \nonumber \\
\nabla_{\theta^b} J^b(\pi_{\theta^b}^b)=\nabla_{\theta^b}\sum_{\Gamma}P(\Gamma|\pi_{\theta^b}^b)R(\Gamma).  \nonumber
\end{align}

They have different trajectories $\Gamma^\star$ and $\Gamma$ because of the different $\pi_\theta$ and $\pi_{\theta^b}^b$. Therefore, if we can remove the impact of the trajectory $\Gamma$ on gradient $\nabla_{\theta^b}\sum_{\Gamma}P(\Gamma|\pi_{\theta^b}^b)$ for POMDP, we can convert the policy gradient theorem in POMDP to Lemma 1. To prove this, we rewrite the police gradient of POMDP as,
\begin{align}
\nabla_{\theta^b} J^b(\pi_{\theta^b}^b) =& \nabla_{\theta^b}\sum_{\Gamma}P(\Gamma|\pi_{\theta^b}^b)R(\Gamma)  \nonumber \\
=& \sum_{\Gamma} \frac{P(\Gamma|\pi_{\theta^b}^b)}{P(\Gamma|\pi_{\theta^b}^b)}\nabla_{\theta^b}P(\Gamma|\pi_{\theta^b}^b)R(\Gamma) \nonumber \\
=& \sum_{\Gamma} P(\Gamma|\pi_{\theta^b}^b) \nabla_{\theta^b}\log P(\Gamma|\pi_{\theta^b}^b)R(\Gamma) \nonumber \\
=& \mathbb{E}_{\Gamma \sim{\pi_{\theta^b}^b}}[\nabla_{\theta^b}\log P(\Gamma|\pi_{\theta^b}^b)R(\Gamma)]. \nonumber 
\end{align}

Therefore, the update direction of the policy is,
\begin{eqnarray}
\nabla_{\theta^b}\log P(\Gamma|\pi_{\theta^b}^b)  \!\!\!\!\! &=&  \!\!\!\!\! \nabla_{\theta^b}\log \bigg{[} P(\bm{s}^b_{0}) P(\bm{o}^b_{0}|\bm{s}^b_{0})   \!\!  \prod_{t_b =1 }^{\mathcal{T}_b} \!\! P(\bm{o}^b_{t_b}|\bm{s}^b_{t_b}) \bigg{.}\nonumber\\
&& \!\!  \prod_{t_b =1 }^{\mathcal{T}_b} \!\! P(\bm{s}^b_{t_b}|\bm{s}^b_{t_b-1},\bm{a}^b_{t_b-1})  \nonumber\\
&& \bigg{.} \!\!  \prod_{t_b =1 }^{\mathcal{T}_b} \!\! \pi_{\theta^b}^b(\bm{a}^b_{t_b-1}|\bm{h}^b_{t_b-1}) \bigg{]}\nonumber \\
&=&  \!\!\!\!\! \nabla_{\theta^b}\log \! \bigg{[} \! P(\bm{s}^b_{0}) \!+ \! P(\bm{o}^b_{0}|\bm{s}^b_{0}) \!+ \!\!  \sum_{t_b =1 }^{\mathcal{T}_b} \!\! P(\bm{o}^b_{t_b}|\bm{s}^b_{t_b}) \bigg{.}\nonumber\\
&& \!\! + \sum_{t_b =1 }^{\mathcal{T}_b} \!\! P(\bm{s}^b_{t_b}|\bm{s}^b_{t_b-1},\bm{a}^b_{t_b-1})  \nonumber\\
&& \bigg{.} \!\!  + \sum_{t_b =1 }^{\mathcal{T}_b} \!\! \pi_{\theta^b}^b(\bm{a}^b_{t_b-1}|\bm{h}^b_{t_b-1}) \bigg{]}\nonumber \\
&=& \!\!\!\!\! \nabla_{\theta^b}\log \! \bigg{[} \!\!  \sum_{t_b =1 }^{\mathcal{T}_b} \!\! \pi_{\theta^b}^b(\bm{a}^b_{t_b-1}|\bm{h}^b_{t_b-1}) \bigg{]}\nonumber \\
&=& \!\!\!\!\! \sum_{t_b =1 }^{\mathcal{T}_b} \! \bigg{[} \nabla_{\theta^b}\log  \pi_{\theta^b}^b(\bm{a}^b_{t_b-1}|\bm{h}^b_{t_b-1}) \bigg{]},\nonumber 
\end{eqnarray}
which is independent of states $\bm{s}^b_{t_b}$. Therefore, we can use $\bm{h}^b_{t_b}$ instead of $\bm{s}^b_{t_b}$, which completes the proof.

\bibliographystyle{IEEEtran}

\bibliography{reference}